\documentclass[twocolumn,showpacs,preprintnumbers,amsmath,amssymb]{revtex4}
\usepackage{epsfig}
\usepackage{dcolumn}% Align table columns on decimal point
\usepackage{bm}% bold math
\usepackage{color}
\newcommand{\beq}{\begin{equation}}
\newcommand{\eeq}{\end{equation}}
\newcommand{\beqa}{\begin{eqnarray}}
\newcommand{\eeqa}{\end{eqnarray}}
\newcommand{\bea}{\begin{array}}
\newcommand{\ea}{\end{array}}

\newcommand{\dd}{\mathrm{d}}

\newcommand{\lag}{\langle}
\newcommand{\rag}{\rangle}

\newcommand{\ii}{{\rm i}}

\newcommand{\vx}{{\bf x}}
\newcommand{\vk}{{\bf k}}
\newcommand{\vq}{{\bf q}}

\newcommand{\tu}{{\tilde{u}}}
\newcommand{\tW}{{\tilde{W}}}

\newcommand{\cO}{{\cal O}}
\newcommand{\cP}{{\cal P}}

\newcommand{\rhob}{\overline{\rho}}

\begin{document}

\title{Accuracy of analytical models of the large-scale matter distribution}

\author{Patrick Valageas}
\affiliation{Institut de Physique Th\'eorique,\\
CEA, IPhT, F-91191 Gif-sur-Yvette, C\'edex, France\\
CNRS, URA 2306, F-91191 Gif-sur-Yvette, C\'edex, France}
\vspace{.2 cm}

\date{\today}
\vspace{.2 cm}

\begin{abstract}

We investigate the possible accuracy that can be reached by analytical
models for the matter density power spectrum and correlation function.
Using a realistic description of
the power spectrum that combines perturbation theory with a halo model,
we study the convergence rate of several perturbative expansion schemes
and the impact of nonperturbative effects, as well as the sensitivity to
phenomenological halo parameters.
We check that the simple reorganization
of the standard perturbative expansion, with a Gaussian damping prefactor,
provides a well-ordered convergence and a finite correlation function that yields
a percent accuracy at the baryon acoustic oscillation peak (as soon as one goes to second order). 
Lagrangian-space expansions
are somewhat more efficient, when truncated at low orders, but may diverge at
high orders. We find that whereas the uncertainty on the halo-profile
mass-concentration relation is not a strong limitation, the uncertainty on the 
halo mass function can severely limit the accuracy of theoretical predictions for
$P(k)$ (this also applies to the power spectra measured in numerical simulations).
The real-space correlation function provides a better separation between
perturbative and nonperturbative effects, which are restricted to
$x \lesssim 10 h^{-1}$Mpc at all redshifts.

\end{abstract}

 \vskip2pc

\maketitle

\section{Introduction}
\label{Introduction}

The growth of large-scale structures in the Universe through gravitational instability
is a key ingredient of modern cosmology \cite{Peebles1980} and an important
probe of cosmological parameters. In particular, future galaxy surveys aim at
a percent precision on a broad range of scales to constrain the dark energy
component \cite{Laureijs2011}. On large scales or at high redshifts, where the
amplitude of the density fluctuations is small, it is sufficient to use linear theory,
whereas on small scales, in the highly nonlinear regime, one must use numerical
simulations or phenomenological models, such as the halo model \cite{Cooray2002},
which are also calibrated on simulations. On intermediate scales, which are the
focus of several observational probes, such as measures of baryon acoustic 
oscillations (BAOs) \cite{Eisenstein1998,Eisenstein2005}, perturbative 
approaches provide
systematic methods to go beyond linear theory and increase the range of
accurate theoretical predictions. 

This has led to a renewed interest in perturbative approaches that go beyond the
standard perturbative expansion \cite{Goroff1986,Bernardeau2002} by
including partial resummations of higher-order terms. A variety of schemes have
been developed, in both the Eulerian-space framework 
\cite{Crocce2006a,Crocce2006b,Valageas2007,Taruya2008,Pietroni2008,Taruya2012,Anselmi2012}
and the Lagrangian-space framework \cite{Matsubara2008,Matsubara2008a,Valageas2013}.
However, most of these approaches are based on the single-stream approximation
and neglect shell-crossing effects (a few exceptions are Refs.~\cite{Pietroni2012,Carrasco2012,Valageas2013}).

To compare these theoretical predictions with observations, it is important to
understand their range of validity. The impact of shell crossing
onto the matter power spectrum has already been investigated in \cite{Afshordi2007}
and \cite{Valageas2011a}, using two variants of a phenomenological halo model
or the Zel'dovich dynamics \cite{Zeldovich1970}, and in \cite{Pueblas2009}
by estimating the generation of vorticity and velocity dispersion.
In this paper, we investigate in more details the convergence of several perturbative
expansions and the quantitative impact of nonperturbative effects due to shell
crossing. Moreover, we estimate the sensitivity of the predicted power spectrum
to the uncertainty of phenomenological parameters (the concentration of the
halo density profile, the halo mass function) that must be taken from numerical
simulations. This also gives an estimate of the accuracy of the power spectra
obtained from these simulations.

To this order, we use the simple analytical model developed in \cite{Valageas2013}, 
which combines one-loop standard perturbation theory with a halo model within
a Lagrangian-space framework.
This provides a good approximation to the nonlinear power spectrum
on a broad range of scales and redshifts while being based on a physical
modeling. Therefore, we can expect that it provides a good quantitative basis for
such a study.
For this article, the advantage of using such a toy model rather than
numerical simulations is that we can easily separate perturbative from
nonperturbative contributions, as well as the impact of different halo parameters.
In contrast, numerical simulations include at once all these effects and this can
lead to misleading comparisons with analytical approaches that neglect some of them.

This paper is organized as follows.
In Sec.~\ref{Toy-model}, we first recall the model for the power spectrum
obtained in \cite{Valageas2013} that is the basis of our study.
Then, in Sec.~\ref{Convergence} we investigate the rate of convergence of
some simple expansion schemes, both within an Eulerian-space and a
Lagrangian-space framework.
Next, in Sec.~\ref{perturbative} we estimate the importance of nonperturbative
contributions, as a function of scale and redshift, and we consider the impact
of the limited accuracy of phenomenological halo parameters.
We conclude in Sec.~\ref{Conclusion}.

\section{Toy model for the fully nonlinear matter power spectrum}
\label{Toy-model}

We briefly recall in this section the expression of the matter power spectrum obtained
in \cite{Valageas2013} (see the Appendix for details).
As in usual halo models \cite{Cooray2002}, this model writes the nonlinear power
spectrum as a sum of one-halo and two-halo terms,
\beq
P(k) = P_{\rm 1H}(k) + P_{\rm 2H}(k) .
\label{Pk-halos}
\eeq
The one-halo term reads as
\beq
P_{\rm 1H}(k) = \int_0^{\infty} \frac{\dd\nu}{\nu} f(\nu) \frac{M}{\rhob (2\pi)^3}
\left(  \tu_M(k) - \tW(k q_M) \right)^2 ,
\label{Pk-1H}
\eeq
where $f(\nu)$ is the scaling function that determines the halo mass function,
as $\dd n= \rhob/M f(\nu) \dd\nu/\nu$, with $\nu=\delta_c/\sigma(M)$
and $\sigma(M)$ the rms linear density contrast at mass scale $M$.
Here, $\tu_M(k)$ is the Fourier transform of the density profile
of a halo of mass $M$ [we use the popular Navarro-Frenk-White (NFW) 
profile \cite{Navarro1997}], given by Eq.(\ref{u-M-def}),
while $\tW(k q_M)$ is the Fourier transform of the top hat of Lagrangian
radius $q_M$ [it also ensures that $P_{\rm 1H}(k) \propto k^4$ at low $k$
in agreement with the conservation of mass and momentum \cite{Peebles1974}].
Using a Lagrangian-space framework, the two-halo term reads as
\beqa
P_{\rm 2H}(k) & = & \int \frac{\dd\vq}{(2\pi)^3} \; F_{\rm 2H}(q) \;
\lag e^{\ii\vk\cdot\vx} \rag^{\rm vir}_{q} \;
\frac{1}{1+A_1} \nonumber \\
&& \hspace{-1.4cm} \times \; e^{-\frac{1}{2} k^2 (1-\mu^2) \sigma_{\perp}^2} \;
\biggl \lbrace e^{-\varphi(-\ii k q \mu \sigma^2_{\kappa})
/\sigma_{\kappa}^2} + A_1 \nonumber \\
&& \hspace{-1.4cm} \!\! + \!\! \int_{0^+-\ii\infty}^{0^++\ii\infty} \frac{\dd y}{2\pi\ii} \; 
e^{-\varphi(y)/\sigma^2_{\kappa}} 
\left( \! \frac{1}{y} - \frac{1}{y\! + \! \ii k q \mu \sigma^2_{\kappa}} \! \right)
\!\! \biggl \rbrace  , \nonumber \\
&& 
\label{Pk-2H-1}
\eeqa
where we integrate over the Lagrangian-space separation $\vq \equiv \vq_2-\vq_1$
of particle pairs and $\mu=\vk\cdot\vq/(k q)$.
Defining the displacement field $\Psi_i=\vx_i-\vq_i$, the longitudinal and transverse
variances (with respect to the direction $\vq$) of the linear relative displacements
are $\sigma_{\parallel}^2=\lag (\Psi_{2\parallel L}-\Psi_{1\parallel L})^2\rag$ and
$\sigma_{\perp}^2=\lag (\Psi_{2\perp L}-\Psi_{1\perp L})^2\rag$ (along any given
transverse direction). They are given by Eqs.(\ref{sig-parallel})-(\ref{sig-perp}).
The dimensionless longitudinal relative displacement is
denoted as $\kappa= x_{\parallel}/q = 1+\Psi_{\parallel}/q$ and
$\sigma_{\kappa}=\sigma_{\parallel}/q$.
As explained in the Appendix, to derive Eq.(\ref{Pk-2H-1}), we approximated the
transverse displacement as Gaussian (as in Lagrangian linear theory) whereas the
longitudinal displacement is non-Gaussian, defined at the perturbative level by its
cumulant generating function $\varphi(y)$,
\beq
\varphi(y) = - \sum_{n=1}^{\infty} \frac{S^{\kappa}_n}{n!} \; (-y)^n , 
\;\;\; S^{\kappa}_n = \frac{\lag\kappa^n\rag_c}{\sigma_{\kappa}^{2(n-1)}} ,
\label{phi-cum}
\eeq
with the behavior at the origin ($S^{\kappa}_1=S^{\kappa}_2=1$)
\beq
y \rightarrow 0 : \;\; \varphi(y) = y - \frac{y^2}{2} + S^{\kappa}_{3} \, 
\frac{y^3}{6} + ...
\label{phi-y-0}
\eeq
This ensures that the underlying probability distribution function
$\cP_{\varphi}(\kappa)$, given by Eq.(\ref{Pkap-par}),
is normalized to unity and obeys the constraint $\lag \Psi\rag=0$, and an adequate
choice of the resummed function $\varphi$ also ensures that $\cP_{\varphi}(\kappa)$
is everywhere positive.
Finally, the factors $F_{2\rm H}$, $\lag e^{\ii\vk\cdot\vx} \rag^{\rm vir}_{q}$,
$A_1$, and the last integral over $y$, are nonperturbative shell-crossing
contributions associated with pancake and halo formation
(see \cite{Valageas2013} for details).

The nonlinear power spectrum (\ref{Pk-halos}) combines
perturbation theory with a halo model. In particular, we showed
in Ref.\cite{Valageas2013} that this power spectrum is exact up to second
order $P_L^2$ if we use for the skewness $S^{\kappa}_{3}$ the expression
(\ref{S3-def}). Then, using for the resummed function $\varphi(y)$
the ansatz (\ref{phi-alpha-def})-(\ref{S3-alpha}), we checked
that we obtained a good agreement with numerical simulations up to
$k \sim 10 h$Mpc$^{-1}$.
Then, we can use the power spectrum (\ref{Pk-halos}) as a toy model to
investigate the rate of convergence of various perturbative expansions to the
resummed perturbative power or to estimate the impact of nonperturbative
contributions and of halo parameters.

\section{Convergence of some perturbative expansions}
\label{Convergence}

\subsection{Eulerian-space expansions}
\label{Eulerian}

Throughout this paper, by ``perturbative'' we refer to quantities that can be
expanded over integer powers of the linear power spectrum $P_L$,
as in the standard perturbation theory \cite{Bernardeau2002}.
Then, the perturbative part of the matter density power spectrum (\ref{Pk-halos})
writes as (see the Appendix and Ref.\cite{Valageas2013})
\beq
P_{\rm pert.}(k) = \int \frac{\dd\vq}{(2\pi)^3} \; 
e^{-\varphi(-\ii k q \mu \sigma_{\kappa}^2)/\sigma_{\kappa}^2} \;
e^{-\frac{1}{2} k^2 (1-\mu^2) \sigma_{\perp}^2} .
\label{P-no-sc-1}
\eeq
It is also the perturbative part of the two-halo component (\ref{Pk-2H-1}), as the
one-halo component (\ref{Pk-1H}) of the form $e^{-1/\sigma^2}$ is nonperturbative.
It is convenient to define the function $\psi(y)$, which describes the deviation
from the Gaussian, by
\beq
\psi(y) \equiv \varphi(y) - y + \frac{y^2}{2} = S^{\kappa}_{3} \, 
\frac{y^3}{6} + ...
\label{psi-y}
\eeq
and the power spectrum (\ref{P-no-sc-1}) writes as
\beqa
P_{\rm pert.}(k) & = & \int \frac{\dd\vq}{(2\pi)^3} \; 
e^{\ii k q \mu -\frac{1}{2} k^2 [\mu^2 \sigma_{\parallel}^2 + (1-\mu^2) \sigma_{\perp}^2]} 
\nonumber \\
&& \times \; e^{-\psi(-\ii k q \mu \sigma_{\kappa}^2)/\sigma_{\kappa}^2} .
\label{P-no-sc-2}
\eeqa
Thus, setting $\psi=0$ in Eq.(\ref{P-no-sc-2}) gives back the Zel'dovich power spectrum
\cite{Zeldovich1970,Schneider1995,Valageas2007a},
which is only exact up to linear order over $P_L$,
whereas keeping $\psi \neq 0$ with Eq.(\ref{S3-def}) gives a power spectrum that is
exact up to second order $P_L^2$ and also generates approximate higher-order
contributions \cite{Valageas2013}.

In this framework, the functions $\varphi(y)$ and $\psi(y)$ depend on the scale $q$,
but they do not depend on redshift nor on the amplitude of the linear power
spectrum $P_L$.
Therefore, the ``standard'' perturbative expansion over powers of $P_L$ 
of the power spectrum (\ref{Pk-halos}) can be
recovered by expanding Eq.(\ref{P-no-sc-2}) over powers of $P_L$, that is, over
powers of the linear displacement variances $\sigma_{\parallel}^2$, $\sigma_{\perp}^2$,
and $\sigma_{\kappa}^2$.
We denote this ``standard perturbation theory'' expansion as
\beq
P_{\rm pert.}(k) = \sum_{n=1}^{\infty} P^{(n)}_{\rm SPT}(k) \;\;\; \mbox{with} \;\;\;
P^{(n)}_{\rm SPT} \propto (P_L)^n ,
\label{SPT-1}
\eeq
and from Eq.(\ref{P-no-sc-2}) each term reads as
\beqa
P^{(n)}_{\rm SPT}(k) & = & \int \frac{\dd\vq}{(2\pi)^3} \; e^{\ii k q \mu} \;
\nonumber \\
&& \hspace{-1.5cm}  \times \;  \lfloor
e^{-\frac{1}{2} k^2 [\mu^2 \sigma_{\parallel}^2 + (1-\mu^2) \sigma_{\perp}^2]
-\psi(-\ii k q \mu \sigma_{\kappa}^2)/\sigma_{\kappa}^2} \rfloor_{(P_L)^n} ,
\label{Pn-SPT-1}
\eeqa
where $\lfloor .. \rfloor_{(P_L)^n}$ denotes the term of order $(P_L)^n$ of the
expression between the two delimiters.
Although the explicit expression (\ref{Pn-SPT-1}) derives from the
Lagrangian-space formulation (\ref{P-no-sc-1}) within our framework,
the standard expansion of the form (\ref{SPT-1}) is usually computed from
a Eulerian-space approach. 
Being uniquely defined as the expansion over powers of $P_L$, the method of
computation does not matter and no trace of the Lagrangian-space framework
remains in this expansion, which can be fully defined within a Eulerian-space
approach.

Because of the approximations involved in the model (\ref{P-no-sc-1}), this
perturbative expansion is only exact up to second order $P_L^2$.
However, we can expect its main features to be correct as the power spectrum
built in Ref.\cite{Valageas2013} has been shown to provide a good quantitative
match to numerical simulations and it is based on a realistic physical modeling
[e.g., the probability distribution function $\cP(\kappa)$ of relative displacements
that underlies Eq.(\ref{P-no-sc-1}) is well behaved].

\begin{figure}
\begin{center}
\epsfxsize=8.5 cm \epsfysize=6.5 cm {\epsfbox{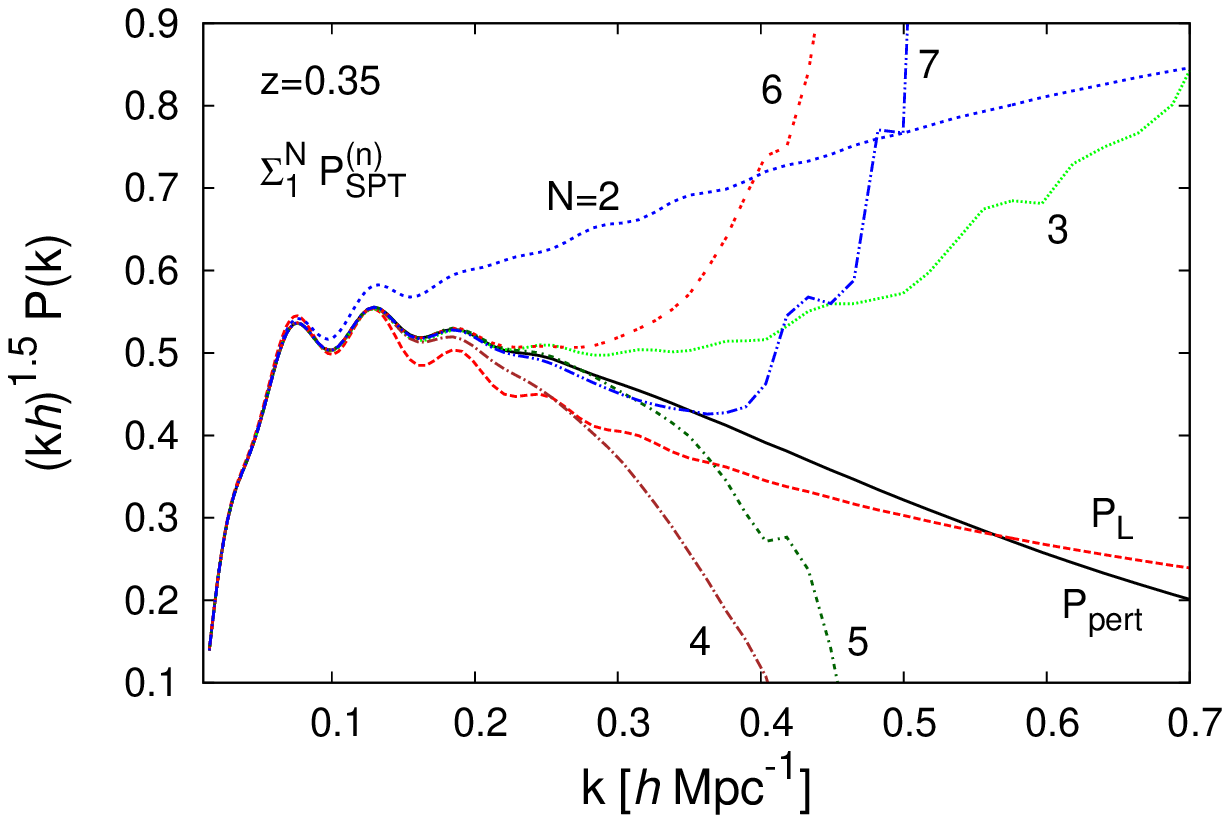}}\\
\epsfxsize=8.5 cm \epsfysize=6.5 cm {\epsfbox{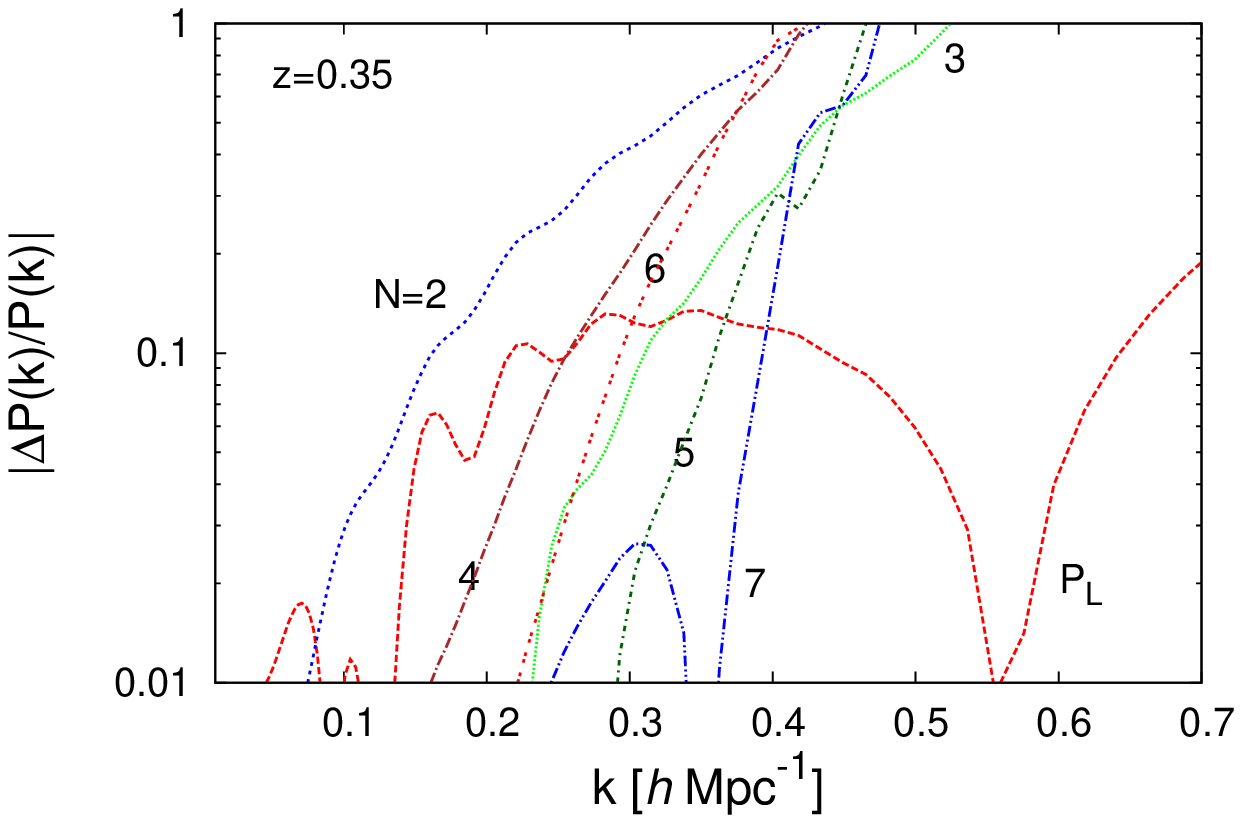}}
\end{center}
\caption{{\it Upper panel:} ``standard'' perturbative expansion over
powers of $P_L$ of the power spectrum (\ref{P-no-sc-1}), as in Eq.(\ref{SPT-1}).
We show the partial series truncated at order $N=1, 2, .., 7$ (the case $N=1$ is
simply the linear power spectrum $P_L$), as well as the resummed perturbative power
spectrum $P_{\rm pert.}$, at redshift $z=0.35$.
{\it Lower panel:} relative deviation between these partial series and the resummed
perturbative power spectrum $P_{\rm pert.}$ of Eq.(\ref{P-no-sc-1}).}
\label{fig-kPk_SPT_z0.35}
\end{figure}

We show the first seven partial series of the expansion (\ref{SPT-1}) in
Fig.~\ref{fig-kPk_SPT_z0.35}.
We recover the well-known behavior of the standard perturbation theory
\cite{Crocce2006a,Valageas2011a},
which has already been exactly computed up to two-loop order, or up to very
high order for the simpler Zel'dovich dynamics.
As seen in the upper panel,  the amplitude of higher orders grows increasingly fast at
high $k$, so that the series (\ref{SPT-1}) is
badly behaved and cannot be used to compute the real-space correlation function
because of the divergent high-$k$ tails. 
However, on quasilinear scales, $k \lesssim 0.4 h$Mpc$^{-1}$ at $z=0.35$, the series
seems to converge, at least up to order $N=7$.
Nevertheless, the convergence is not very regular, as the series truncated
at orders $N=2,4,$ or $6$, shows a stronger deviation from the full perturbative power
(\ref{P-no-sc-2}) than the previous orders $N=1,3,$ or $5$ (except on the very
large scales).  
This faster convergence of odd-order partial series is even more clearly seen in the
lower panel.
This is due to the change of signs of the
fast growing contributions $P^{(n)}_{\rm SPT}$. The nonlinear power spectrum
(in the convergence domain) arises from cancellations between the different terms
$P^{(n)}$ and this explains why, for some values of $N$ going to order $N+1$ can
worsen the result on scales that have not converged yet, because some required
counterterms are included in the subsequent order $N+2$.
In particular, while going to third order over $P_L$ (i.e., two-loop order in terms of the
usual perturbative diagrams) significantly extends the range of validity of the prediction
as compared with the linear or second-order approximations, a better approximation
requires going to fifth order.

On the other hand, if we compare the partial series (\ref{SPT-1}) with the full nonlinear
power spectrum
$P(k)$ measured in numerical simulations, or given by Eq.(\ref{Pk-halos}), we
find that the second-order (i.e., one-loop) approximation fares best than all other
truncations on a broad
range of scale. This is because $k^{1.5}P_{\rm 1-loop}(k)$ happens to show a slow
rise beyond quasilinear wave numbers that is similar to the growth shown by the
nonlinear power spectrum. However, this is only a misleading coincidence: the
``true'' perturbative power spectrum $P_{\rm pert.}$, of which $P_{\rm 1-loop}(k)$ is only
a second-order approximation, actually shows a faster decrease at high $k$ and
the growth of the nonlinear power spectrum $P(k)$ is due to nonperturbative effects
that are not included in any perturbative scheme based on the single-stream
approximation.
This emphasizes the danger of comparing various perturbative approaches (or more
general analytic models) with numerical simulations, which do not separate between
the different contributions to the power spectrum (e.g., originating from perturbative
and nonperturbative scales). Thus, a seemingly good agreement between a
perturbative prediction and the full nonlinear power spectrum on transition scales
is not necessarily meaningful. Because a non-negligible part of the power comes from
effects that are not included in the model, a good match is likely to be a coincidence
rather than the result of a very realistic and accurate modeling, and it may even become
a problem as one tries to improve the model by adding these other effects.

\begin{figure}
\begin{center}
\epsfxsize=8.5 cm \epsfysize=6.5 cm {\epsfbox{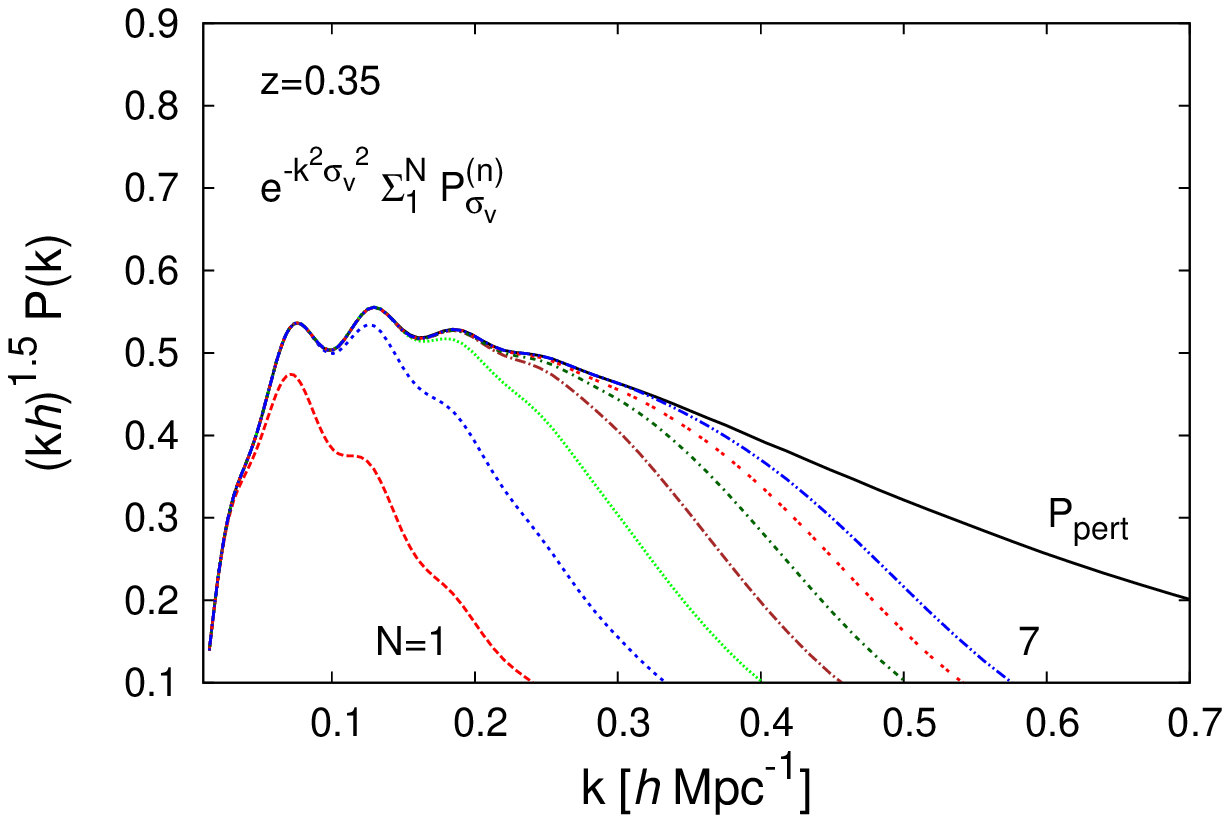}}\\
\epsfxsize=8.5 cm \epsfysize=6.5 cm {\epsfbox{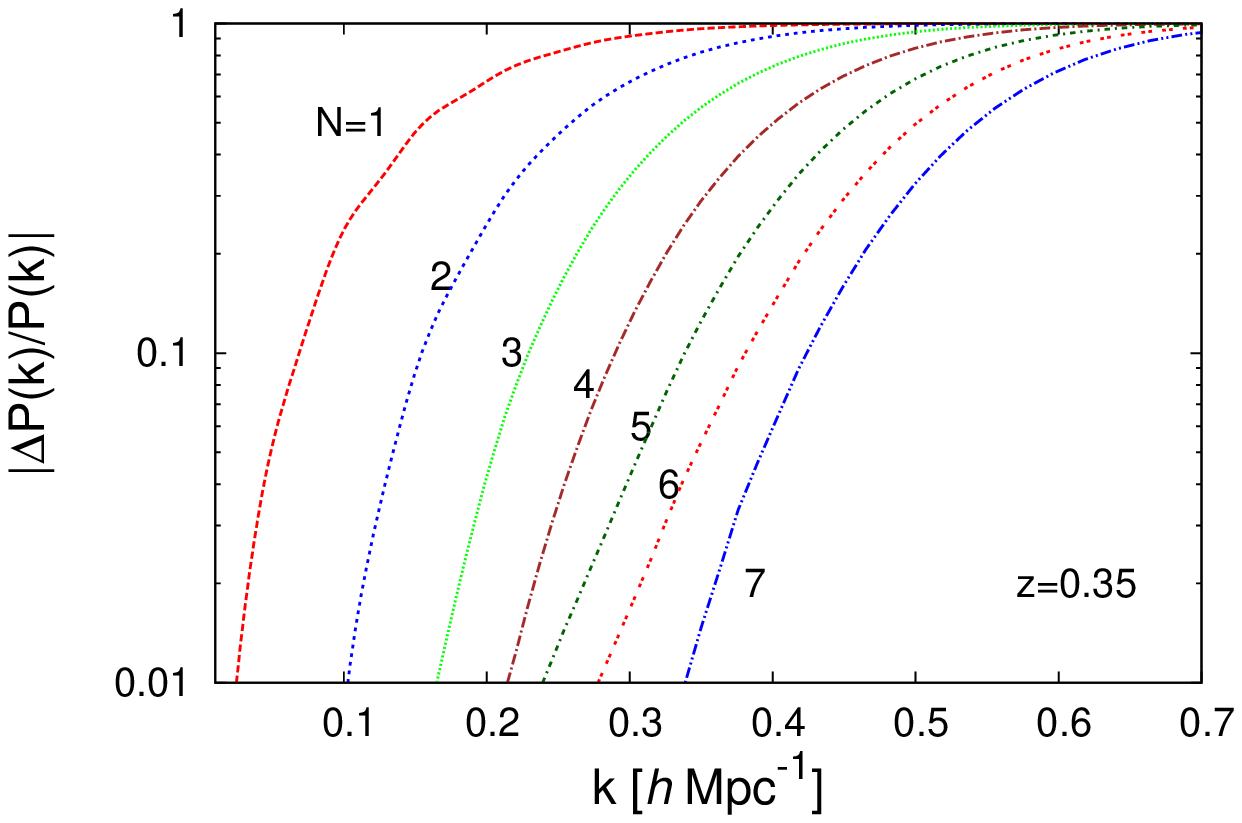}}
\end{center}
\caption{{\it Upper panel:} reorganized perturbative expansion (\ref{sigv-1})
of the power spectrum (\ref{P-no-sc-1}), with the Gaussian prefactor
$e^{-k^2 \sigma_v^2}$. 
We show the partial series truncated at order $N=1, 2, .., 7$, as well as the resummed
perturbative power spectrum $P_{\rm pert.}$.
{\it Lower panel:} relative deviation between these partial series and the resummed
perturbative power spectrum $P_{\rm pert.}$ of Eq.(\ref{P-no-sc-1}).}
\label{fig-kPk_sig_z0.35}
\end{figure}

As advocated in \cite{Crocce2006a,Crocce2006b}, it is possible to reorganize
the standard perturbation theory by factoring out a Gaussian damping term
$e^{-k^2 \sigma_v^2}$, where $\sigma_v^2=\lag | \Psi_i |^2 \rag/3$ is the variance
of the linear one-point displacement along one dimension.
We denote this expansion as
\beq
P_{\rm pert.}(k) = e^{-k^2\sigma_v^2} \sum_{n=1}^{\infty} P^{(n)}_{\sigma_v}(k) \;\;\; 
\mbox{with} \;\;\; P^{(n)}_{\sigma_v} \propto (P_L)^n ,
\label{sigv-1}
\eeq
and from Eq.(\ref{P-no-sc-2}) each term reads as
\beqa
\hspace{-0.5cm} P^{(n)}_{\sigma_v}(k) & = & \int \frac{\dd\vq}{(2\pi)^3} \; 
e^{\ii k q \mu} \; \nonumber \\
&& \hspace{-1.5cm}  \times \;  \lfloor e^{-\frac{1}{2} k^2 
[\mu^2 \sigma_{\parallel}^2 + (1-\mu^2) \sigma_{\perp}^2-2\sigma_v^2]
- \psi(-\ii k q \mu \sigma_{\kappa}^2)/\sigma_{\kappa}^2} \rfloor_{(P_L)^n} .
\label{Pn-sigv-1}
\eeqa
The prefactor $e^{-k^2 \sigma_v^2}$ in Eq.(\ref{sigv-1}) arises from the
large-distance limit of the Gaussian term in Eq.(\ref{P-no-sc-2}).
Indeed, at large separation length, $q\rightarrow \infty$, the two particles become
uncorrelated and $\sigma_{\parallel}^2$ and $\sigma_{\perp}^2$ converge to
$2\sigma_v^2$. (This also means that for large $q$ the Gaussian term
in Eq.(\ref{Pn-sigv-1}) goes to zero, which simplifies the numerical computation.)
Again, although the explicit expression (\ref{Pn-sigv-1}) derives from a
Lagrangian-space formulation, the expansion (\ref{sigv-1}) is usually computed
from a Eulerian-space approach and does not require introducing a
Lagrangian-space framework.

The two expansions (\ref{SPT-1}) and (\ref{sigv-1}) can be derived from each other
for any truncation order $N$. For instance, from the definitions
(\ref{SPT-1}) and (\ref{sigv-1}) we obtain at once
\beq
P^{(n)}_{\rm SPT}(k) = \sum_{p=0}^{n-1} \frac{(-k^2\sigma_v^2)^p}{p!} \;
 P^{(n-p)}_{\sigma_v}(k) .
\eeq
(This is in fact how we computed the standard high-order terms $P^{(n)}_{\rm SPT}$
because the terms $P^{(n)}_{\sigma_v}$ are better behaved.)

We show the first seven partial series of the expansion (\ref{Pn-sigv-1}) in
Fig.~\ref{fig-kPk_sig_z0.35}. We recover the well-known property
\cite{Crocce2006a,Valageas2011a} that this reorganized
expansion is much better behaved than the standard expansion (\ref{SPT-1}).
This is partly artificial as this is mostly due to the too strong Gaussian cutoff
$e^{-k^2 \sigma_v^2}$. As explained for instance in \cite{Valageas2007a,Valageas2008},
this damping only occurs for different-time propagators or power spectra and vanishes
for equal-time statistics, which show a smoother power-law decline at high $k$ (as for
the Zel'dovich power spectrum \cite{Schneider1995,Taylor1996,Valageas2007a}).
Then, the series in Eq.(\ref{sigv-1}) must compensate for this too strong cutoff
and behave as $e^{k^2 \sigma_v^2}$, up to power-law corrections, which gives
contributions of the form
$e^{-k^2 \sigma_v^2} P^{(n)}_{\sigma_v}(k) \sim e^{-k^2 \sigma_v^2} (k\sigma_v)^{2n}/n!$
that are positive and well ordered, with a sharp peak around
$k_n \sim \sqrt{n}/\sigma_v$.
Nevertheless, this reorganization of the perturbative expansion provides a very regular
convergence to the perturbative power spectrum (\ref{P-no-sc-1}), at least up to order
$N=7$, The comparison of Fig.~\ref{fig-kPk_sig_z0.35} with the standard expansion
displayed in Fig.~\ref{fig-kPk_SPT_z0.35} shows that even-order series ($N=2,4,6$) are
significantly improved while odd-order series ($N=1,3,5,7$) fare somewhat worse.
However, the well-ordered convergence of the expansion (\ref{sigv-1})
(at least on these scales, as the radius of convergence of the perturbative series is not
necessarily infinite) makes it superior to the standard expansion. Another key
advantage of the series (\ref{sigv-1}) is that the high-$k$ tail is no longer divergent.
This means that we can now compute the Fourier transform of the power spectrum
(\ref{sigv-1}), which gives a perturbative expansion of the two-point correlation
$\xi(x)$.

\begin{figure}
\begin{center}
\epsfxsize=8.5 cm \epsfysize=6.5 cm {\epsfbox{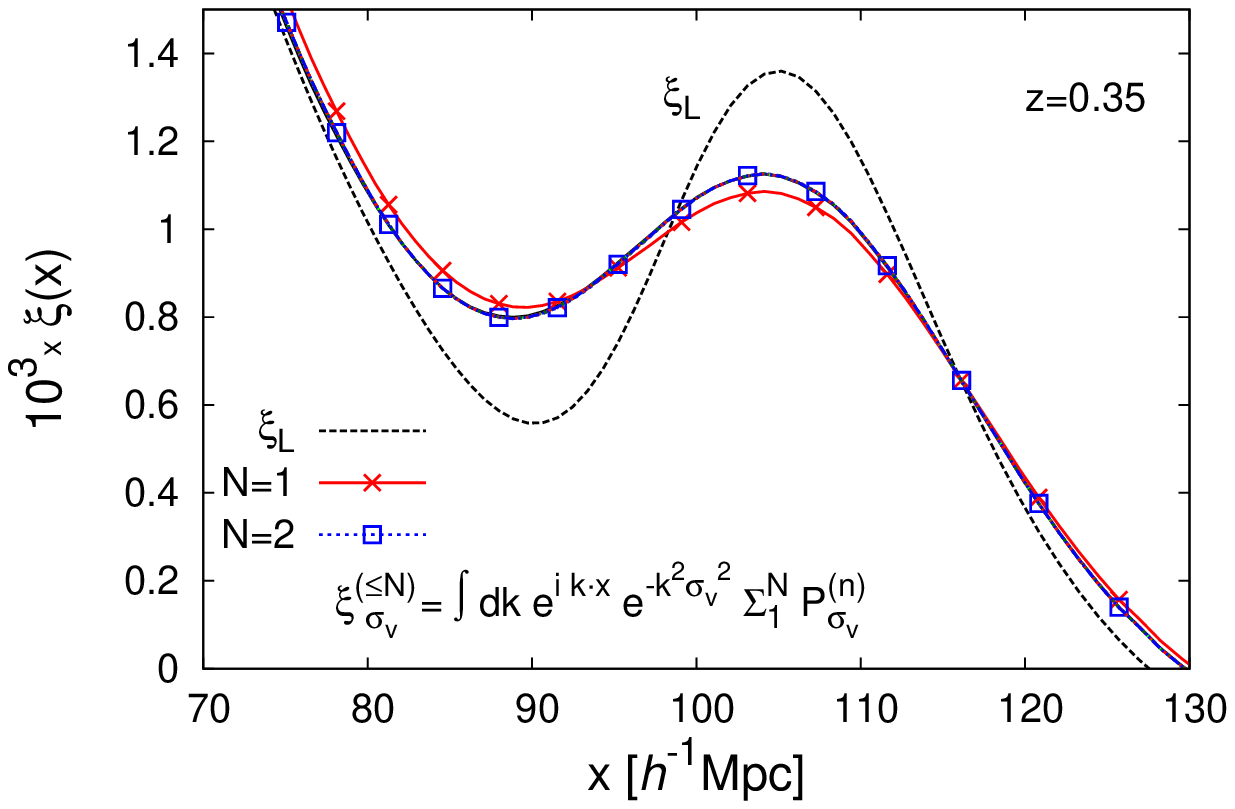}}\\
\epsfxsize=8.5 cm \epsfysize=6.5 cm {\epsfbox{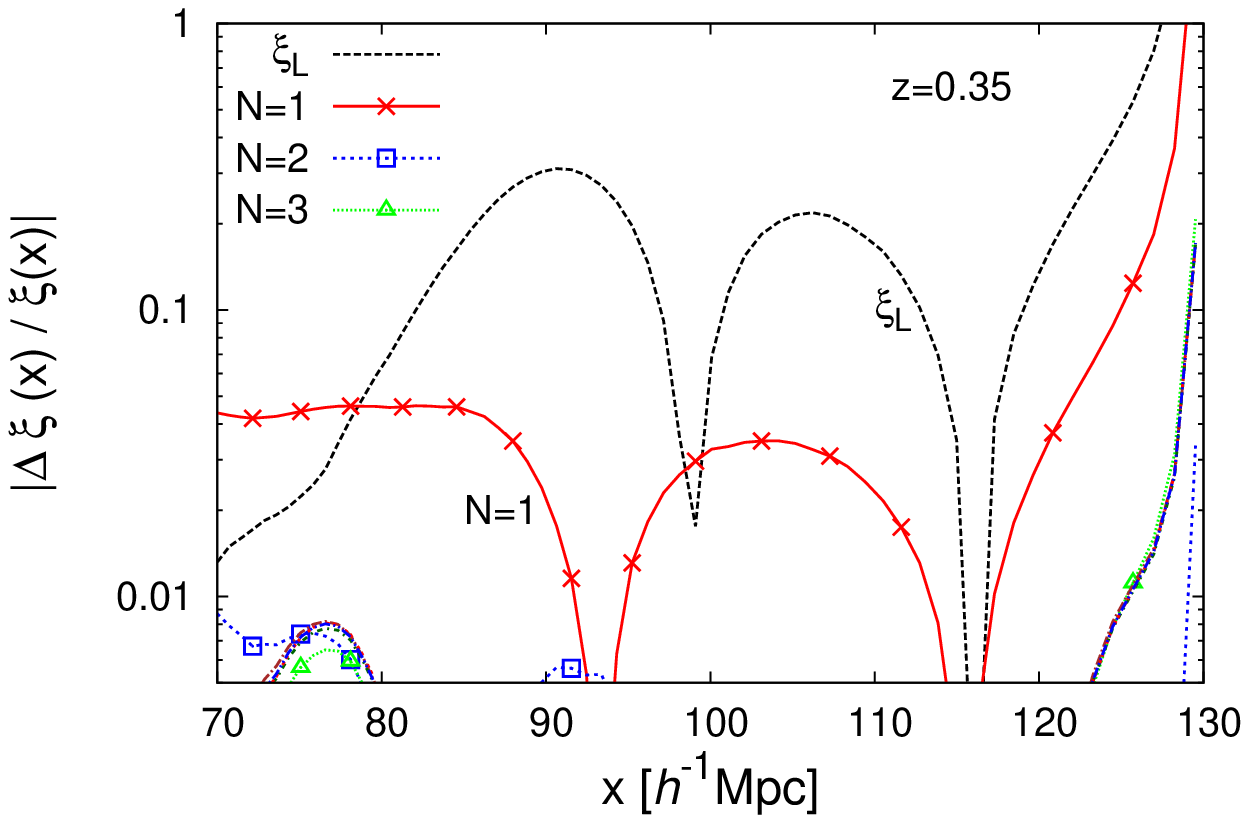}}
\end{center}
\caption{{\it Upper panel:} two-point correlation functions $\xi^{(\leq N)}_{\sigma_v}(x)$
defined by the partial series for the power spectrum shown in
Fig.~\ref{fig-kPk_sig_z0.35} and given by Eq.(\ref{sigv-1}).
We show the partial series truncated at order $N=1, 2, .., 7$, as well as the
resummed correlation $\xi_{\rm pert.}$ and the linear correlation $\xi_L$.
The curves for $N\geq 2$ and $\xi_{\rm pert.}$ cannot be distinguished.
{\it Lower panel:} relative deviation between these perturbative expansions
$\xi^{(\leq N)}_{\sigma_v}$ and the resummed correlation $\xi_{\rm pert.}$.}
\label{fig-xi_sig_z0.35}
\end{figure}

We show in Fig.~\ref{fig-xi_sig_z0.35} the correlation functions $\xi^{(\leq N)}_{\sigma_v}$
obtained from the partial series (\ref{sigv-1}) truncated at order $N$.
We compare these results with the correlation function $\xi_{\rm pert.}$ defined
by the resummed perturbative power spectrum of Eq.(\ref{P-no-sc-1}).
We focus on BAO scales because small nonlinear
scales are beyond the reach of this approach.
As compared with linear theory, we can see that the simple multiplication of the
linear power spectrum by the Gaussian damping $e^{-k^2 \sigma_v^2}$ already provides
a very significant improvement, as the deviation from the resummed correlation
$\xi_{\rm pert.}$ decreases from $30\%$ to $4\%$, at redshift $z=0.35$.
The second (one-loop) order already gives a better than percent accuracy.
(The rise of the curves in the lower panel at $x \sim 130 h^{-1}$Mpc is due to the
change of sign and crossing through zero of $\xi_{\rm pert.}$, which amplifies relative
deviations.)
Thus, perturbative expansions converge very fast for the real-space BAO peak,
provided the high-$k$ tail of their power spectrum is well behaved.
This explains why most perturbative resummation schemes manage to give accurate
predictions for the BAO correlation function. 

The comparison between Figs.~\ref{fig-kPk_sig_z0.35} and
\ref{fig-xi_sig_z0.35} shows that the convergence is much faster for the correlation
function than for the power spectrum.
This means that the real-space BAO peak is a more robust probe of cosmology than
the oscillations of the power spectrum. This is because by looking at the BAO
peak we focus on a fixed scale, far in the quasilinear regime, whereas by looking
at oscillations in $P(k)$ we must consider a broad range of wave numbers, where
different orders contribute, including nonperturbative effects that are not shown
in Fig.~\ref{fig-kPk_sig_z0.35}.
More generally, the power spectrum and the correlation function are not identical
probes for practical purposes, because the Fourier transform mixes all scales, to some
degree, and we can never observe all scales nor make accurate predictions for all scales.

\subsection{Lagrangian-space expansions}
\label{Lagrangian}

We have described in Sec.~\ref{Eulerian} the two simplest perturbative
expansions of the density power spectrum. Being defined as the truncation
at order $N$ of expansions over powers of $P_L$, they can be computed
by any perturbative method, using either a Eulerian or Lagrangian framework.
In practice, they are computed using the standard Eulerian perturbation theory
\cite{Bernardeau2002}, which is the simplest approach and directly provides these
partial series.
However, because our model (\ref{P-no-sc-1}) is based on a Lagrangian-space
framework, it also allows us to investigate the standard Lagrangian perturbation
theory. In this approach, instead of looking for a perturbative expansion of the
density and velocity fields, which gives in turn the density and velocity power spectra,
one looks for a perturbative expansion of the displacement field, $\Psi_i=\vx_i-\vq_i$.
This gives in turn the density power spectrum through the relation
\beqa
P(k) & = & \int\frac{\dd\vq}{(2\pi)^3} \, \lag e^{\ii \vk \cdot \vx} \rag 
\label{Pk-def} \\
& = & \int\frac{\dd\vq}{(2\pi)^3} \, \exp \left[ \sum_{n=1}^{\infty} 
\frac{\lag (\ii \vk \cdot \vx)^n \rag_c}{n!} \right] ,
\label{Pk-cum}
\eeqa
where again $\vq=\vq_2-\vq_1$ and $\vx=\vx_2-\vx_1$ are the Lagrangian and
Eulerian pair separations (and we discarded a Dirac factor).
If we expand the exponential over powers of $P_L$ we recover the standard
Eulerian perturbation theory, as in Sec.~\ref{Eulerian}, but by keeping some
terms in the exponential we obtain alternative approximations, which can be
seen as partial resummations of the standard Eulerian perturbation theory.

If we truncate the cumulant series in the exponential in Eq.(\ref{Pk-cum}) at order
$P_L$, which corresponds to the linear displacement field, we recover the
Zel'dovich power spectrum,
\beq
P_{\rm Z}(k) = \int\frac{\dd\vq}{(2\pi)^3} \; e^{\ii k q \mu 
-\frac{1}{2}k^2[\mu^2\sigma_{\parallel}^2+(1-\mu^2)\sigma_{\perp}^2]} ,
\label{PZ-def}
\eeq
which coincides with Eq.(\ref{P-no-sc-2}) where we set $\psi=0$.
This is only exact up to linear order over $P_L$.
Then, to go beyond the Zel'dovich approximation, we usually compute the displacement
field up to some finite order over the amplitude of the initial fluctuations $\delta_L$,
substitute into Eq.(\ref{Pk-def}) and expand the exponential over terms that are
cubic or higher order over $\delta_L$. This provides a Gaussian expression over
$\delta_L$, with polynomial prefactors, that can be explicitly computed.
Within our framework (\ref{P-no-sc-1}), this simply corresponds to expanding the
last term in Eq.(\ref{P-no-sc-2}) over powers of $P_L$.
This gives the perturbative expansion
\beq
P_{\rm pert.}(k) = \sum_{n=1}^{\infty} P^{(n)}_{\rm Z}(k) \;\;\; \mbox{with} \;\;\;
P^{(n)}_{\rm Z} \sim \cO[(P_L)^n] ,
\label{Z-1}
\eeq
and
\beqa
n \geq 2 : \;\;\; P^{(n)}_{\rm Z}(k) & = & \int \frac{\dd\vq}{(2\pi)^3} \; 
e^{\ii k q \mu-\frac{1}{2} k^2 [\mu^2 \sigma_{\parallel}^2 + (1-\mu^2) \sigma_{\perp}^2]} 
\nonumber \\
&&  \times \;  \lfloor e^{- \psi(-\ii k q \mu \sigma_{\kappa}^2)/\sigma_{\kappa}^2} \rfloor_{(P_L)^n} ,
\label{Pn-Z-1}
\eeqa
and we define $P^{(1)}_{\rm Z}=P_{\rm Z}$.
Thus, each term $P^{(n)}_{\rm Z}$ scales as $(P_L)^n$ for $P_L\rightarrow 0$ but it
also contains contributions at all higher orders.
The first order $n=1$ is the Zel'dovich power spectrum. The second order $n=2$
involves the skewness $S_3^{\kappa}$.

\begin{figure}
\begin{center}
\epsfxsize=8.5 cm \epsfysize=6.5 cm {\epsfbox{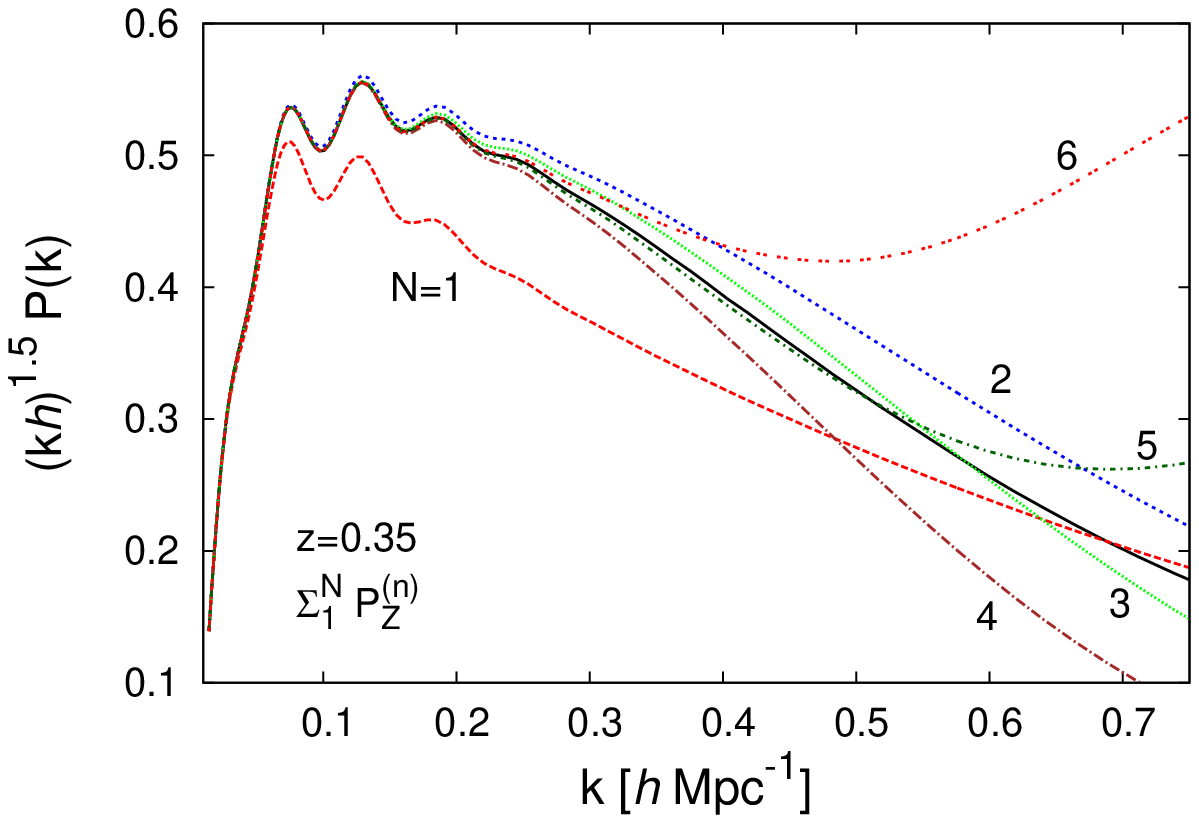}}\\
\epsfxsize=8.5 cm \epsfysize=6.5 cm {\epsfbox{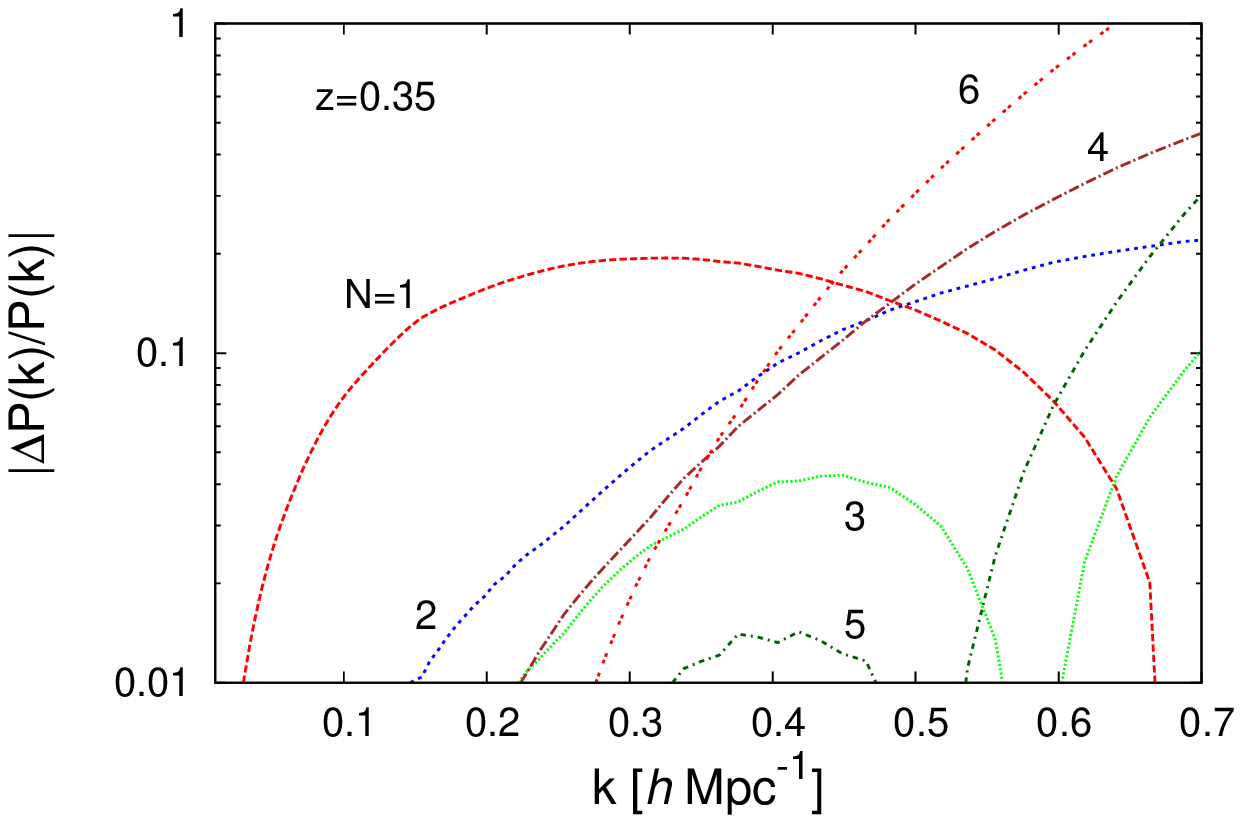}}
\end{center}
\caption{{\it Upper panel:} Lagrangian-space perturbative expansion (\ref{Z-1})
of the power spectrum (\ref{P-no-sc-2}). 
We show the partial series truncated at order $N=1, 2, .., 6$, as well as the resummed
perturbative power spectrum $P_{\rm pert.}$.
{\it Lower panel:} relative deviation between these partial series and the resummed
perturbative power spectrum $P_{\rm pert.}$ of Eq.(\ref{P-no-sc-2}).}
\label{fig-kPk_Z_z0.35}
\end{figure}

We show the first six partial series of the expansion (\ref{Z-1}) in
Fig.~\ref{fig-kPk_Z_z0.35}. As compared with the Eulerian expansions, we can
see that the first few partial series converge much faster and on a broader range
of scales. Moreover, the accuracy improves with the order from $N=1$ to $N=3$,
but the orders $N=4$ and $N=6$ are somewhat worse than $N=3$ and $N=5$.
This suggests that this Lagrangian-space perturbative expansion has a better
start because the Zel'dovich approximation is a good starting point (which is sometimes
used for instance to initialize numerical simulations), as it corresponds to a physical
matter distribution that is realistic on large scales and well defined beyond shell crossing.
In particular, the high-$k$ tail remains well behaved until $N \leq 4$. However, for
high orders, $N=5,6$, the series does not seem to converge very well and the high-$k$
tail starts diverging.
Therefore, the Lagrangian-space expansion (\ref{Z-1}) shares some features with both
Eulerian-space expansions (\ref{SPT-1}) and (\ref{sigv-1}). The first few orders show
an ordered systematic convergence, as for (\ref{sigv-1}), but at higher orders the
convergence becomes irregular, as in (\ref{SPT-1}), and may even break down at high
$k$.
Nevertheless, this Lagrangian-space expansion appears superior to both Eulerian-space
expansions in the sense that for a given low order (e.g., $N=3$ or $5$), the accuracy
is significantly better and has a broader range of validity.
However, it seems that it should not be pushed too far.
The bad convergence at high $k$ and large orders may be due to a finite radius of
convergence of perturbative expansions of the power spectrum (\ref{P-no-sc-1}),
associated with singularities of the function $\varphi(y)$ in the complex plane
at a finite distance from the origin.
Then, the Eulerian-space expansions (\ref{SPT-1}) and (\ref{sigv-1}) are expected
to show the same problems as we push them to higher orders and higher $k$.
On the other, this high-order behavior may not generalize to the exact gravitational
dynamics, where the conditions of convergence of perturbative expansions are not
known.

\begin{figure}
\begin{center}
\epsfxsize=8.5 cm \epsfysize=6.5 cm {\epsfbox{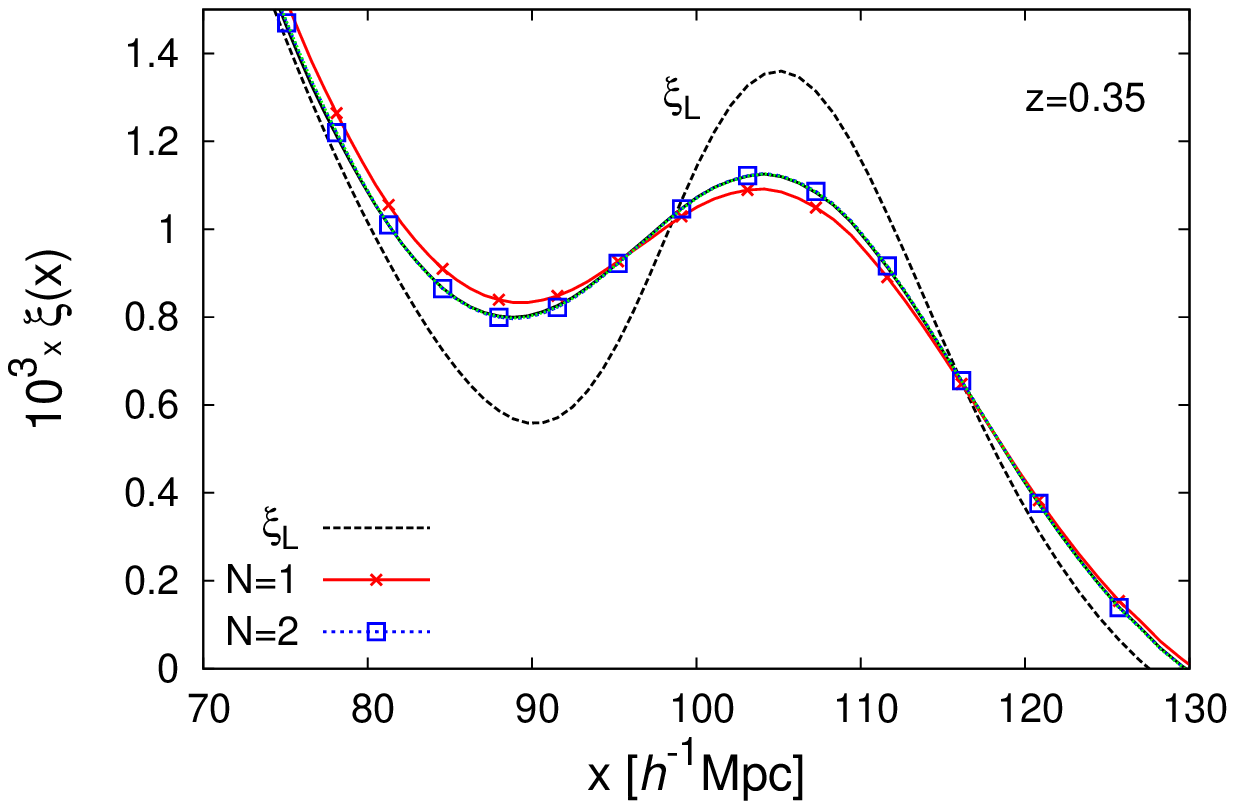}}\\
\epsfxsize=8.5 cm \epsfysize=6.5 cm {\epsfbox{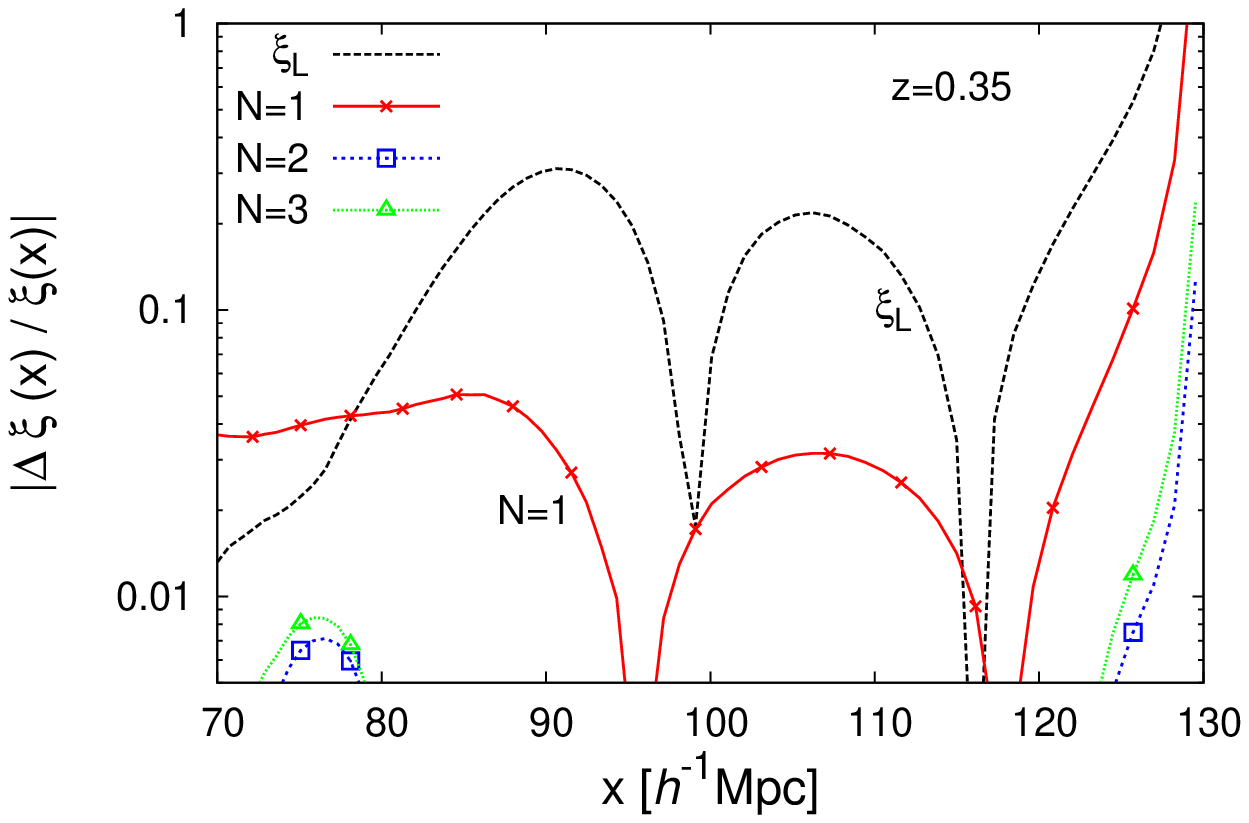}}
\end{center}
\caption{{\it Upper panel:} two-point correlation functions $\xi^{(\leq N)}_{\rm Z}(x)$
defined by the partial series for the power spectrum shown in
Fig.~\ref{fig-kPk_Z_z0.35} and given by Eq.(\ref{Z-1}).
We show the partial series truncated at order $N=1, 2$, and $3$, as well as the
resummed correlation $\xi_{\rm pert.}$ and the linear correlation $\xi_L$.
The curves for $N=2,3$, and $\xi_{\rm pert.}$, cannot be distinguished.
{\it Lower panel:} relative deviation between these perturbative expansions
$\xi^{(\leq N)}_{\sigma_v}$ and the resummed correlation $\xi_{\rm pert.}$.}
\label{fig-xi_Z_z0.35}
\end{figure}

We show the correlation functions $\xi^{(\leq N)}_{\rm Z}$ obtained from the
partial series (\ref{Z-1}) in Fig.~\ref{fig-xi_Z_z0.35}
We again focus on BAO scales, for the orders $N=1,2$, and $3$.
As for the reorganized Eulerian-space expansion (\ref{sigv-1}) shown in
Fig.~\ref{fig-xi_sig_z0.35}, the first term $N=1$ (which is the Zel'dovich approximation
here) already improves the accuracy from $30\%$ to $4\%$ at redshift $z=0.35$, as
compared with linear theory. 
This reasonably good agreement of the Zel'dovich correlation function with numerical
simulations on large scales was already noticed in \cite{Valageas2013,Carlson2013}.
Orders $N=2$ and $3$ provide a subpercent accuracy.
The accuracy is only slightly better than for the expansion (\ref{sigv-1}) at these
orders.

\begin{figure}
\begin{center}
\epsfxsize=8.5 cm \epsfysize=6.5 cm {\epsfbox{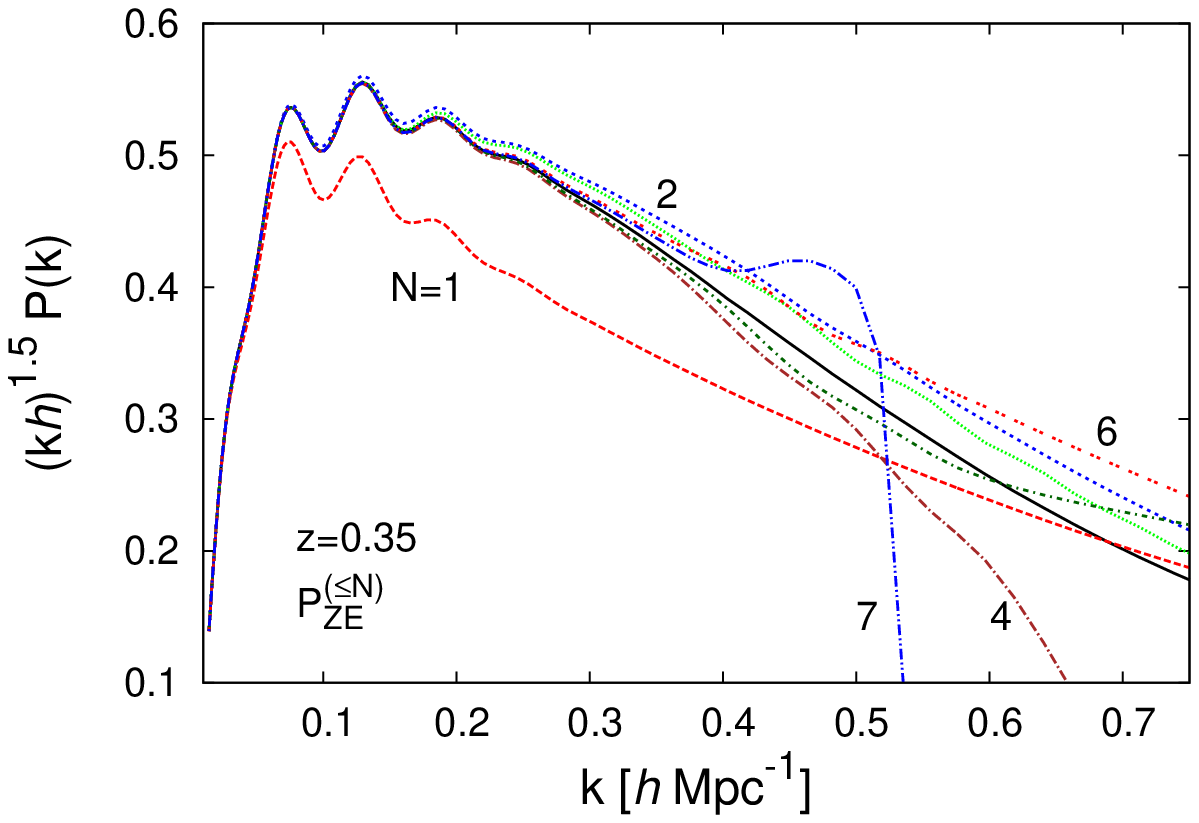}}\\
\epsfxsize=8.5 cm \epsfysize=6.5 cm {\epsfbox{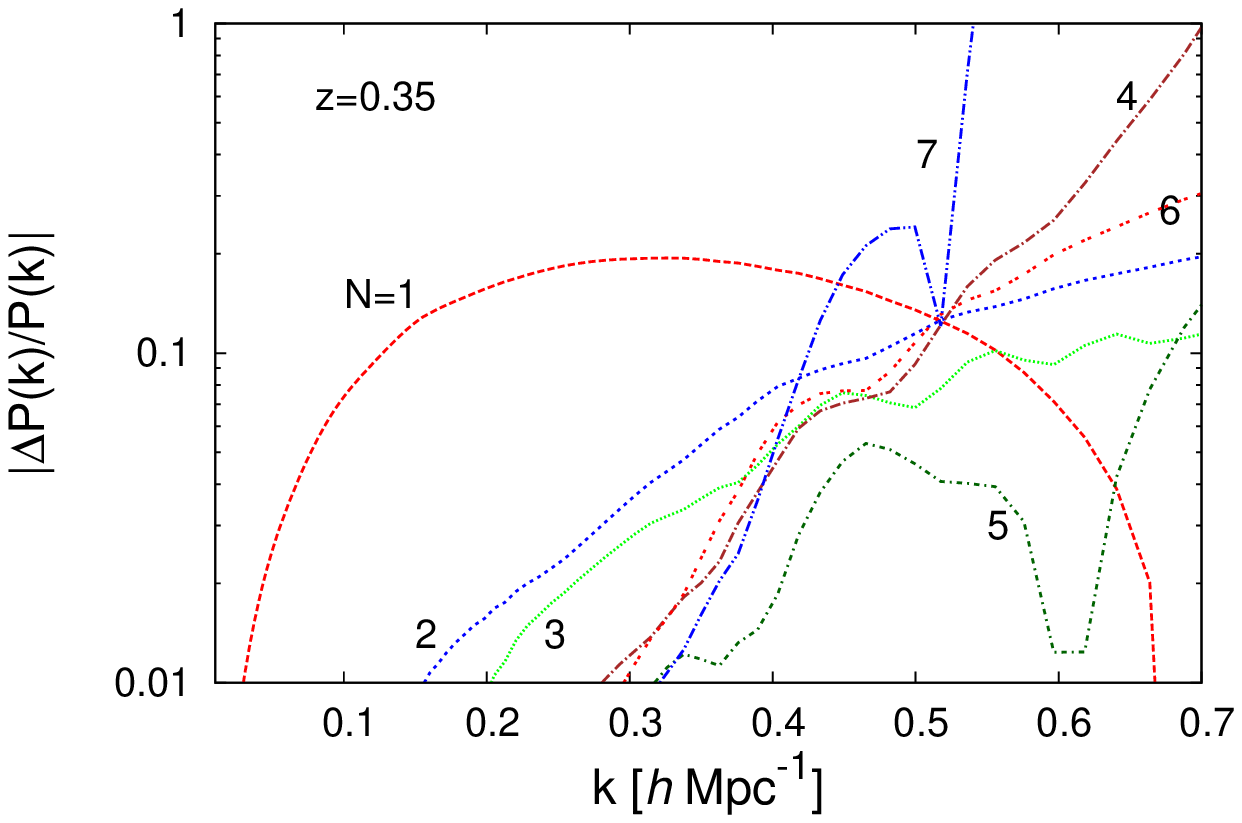}}
\end{center}
\caption{{\it Upper panel:} Lagrangian-space perturbative expansion (\ref{ZE-1})
of the power spectrum (\ref{P-no-sc-2}). 
We show the orders $N=1, 2, .., 7$, as well as the resummed
perturbative power spectrum $P_{\rm pert.}$.
{\it Lower panel:} relative deviation between these partial series and the resummed
perturbative power spectrum $P_{\rm pert.}$ of Eq.(\ref{P-no-sc-2}).}
\label{fig-kPk_ZE_z0.35}
\end{figure}

An alternative to the approach associated with the expansion (\ref{Pn-Z-1}) is to
again compute the displacement field up to some finite order over the initial
fluctuations $\delta_L$, but then to substitute into Eq.(\ref{Pk-cum}) and keep
a truncated cumulant series into the exponential.
Within our framework (\ref{P-no-sc-2}), this simply corresponds to expanding the
function $\psi$ in the exponential.
This gives the sequence of approximations
\beqa
P_{\rm ZE}^{(\leq N)}(k)  & = & \int \frac{\dd\vq}{(2\pi)^3} \; 
e^{\ii k q \mu-\frac{1}{2} k^2 [\mu^2 \sigma_{\parallel}^2 + (1-\mu^2) \sigma_{\perp}^2]} 
\nonumber \\
&&  \times \;  e^{\lfloor - \psi(-\ii k q \mu \sigma_{\kappa}^2)/\sigma_{\kappa}^2 \rfloor_{\leq (P_L)^N} } ,
\label{ZE-1}
\eeqa
where $\lfloor ..  \rfloor_{\leq (P_L)^N}$ denotes the truncation at order $N$ over $P_L$
of the expression between the two delimiters.
The ``E'' in the subscript ``ZE'' recalls that we keep the terms in the exponential.
The order $N=1$ again corresponds to the Zel'dovich power spectrum
(\ref{PZ-def}).

We show the first seven orders of the expansion (\ref{ZE-1}) in
Fig.~\ref{fig-kPk_ZE_z0.35}. We obtain a behavior that is similar to the one
obtained in Fig.~\ref{fig-kPk_Z_z0.35} for the other Lagrangian-space expansion
(\ref{Z-1}).
We again find a fast convergence of the first few orders at $k \leq 0.5 h$Mpc$^{-1}$
and signs of bad behavior and divergence at higher orders and higher $k$.
The accuracy is typically of the same order as for the expansion (\ref{Z-1}).
This means that on quasilinear scales there is not much difference between
expanding the exponential or not. The high-$k$ tail is more sensitive to the details
of the method but it also seems beyond the scope of these perturbative expansions.

We also computed the two-point correlations $\xi^{(\leq N)}_{\rm ZE}$ for $N=1$ and
$2$ and found results that are similar to those in Fig.~\ref{fig-xi_Z_z0.35}.

Thus, we can conclude that for a fixed low order of truncation, Lagrangian-space
approaches can have a broader range of validity than their Eulerian counterparts
for the power spectrum and also provide good approximations to the
two-point correlation on BAO scales.
However, for the computation of the BAO peak of the two-point correlation, the simple 
reorganization (\ref{sigv-1}) of standard Eulerian perturbation theory may be more 
efficient because of its greater simplicity.

\section{Scope of perturbative approaches}
\label{perturbative}

\subsection{Impact of nonperturbative contributions}
\label{Impact-of-nonperturbative-contributions}

We have investigated in Sec.~\ref{Convergence} the convergence of Eulerian
and Lagrangian perturbative expansions toward the resummed perturbative
power spectrum (\ref{P-no-sc-1}).
However, even if we find an efficient perturbative scheme, or manage to resum the
perturbative expansion, this is not sufficient to provide the matter density power
spectrum. Indeed, these perturbative expansions are restricted to the single-stream
regime and do not include nonperturbative effects associated with shell crossings.
(Even though the Lagrangian expansions of Sec.~\ref{Lagrangian} implicitly
go beyond shell crossing, as in the Zel'dovich approximation, this analytic continuation
is not exact and cannot be trusted in this regime.)

Our framework (\ref{Pk-halos})-(\ref{Pk-2H-1}), based on the halo model, combines
these perturbative single-stream contributions with nonperturbative contributions
that are modeled in a more phenomenological fashion. However, because our
model has been shown to provide a good agreement with numerical simulations for
a variety of cosmologies \cite{Valageas2013} and it is based on a realistic
modelization, it allows us to estimate the relative importance of these contributions.
We can split the full nonlinear power spectrum into three components,
\beq
P(k) = P_{\rm pert.}(k) + P_{\rm 2H}^{\rm nonpert.}(k) + P_{\rm 1H}(k) .
\label{P-3split}
\eeq
The one-halo term $P_{\rm 1H}$ is the fully nonperturbative contribution given by
Eq.(\ref{Pk-1H}) and we split the two-halo term given by Eq.(\ref{Pk-2H-1})
into its perturbative part $P_{\rm pert.}$, given by Eq.(\ref{P-no-sc-1}), and its
nonperturbative part $P_{\rm 2H}^{\rm nonpert.}$, defined as
\beq
P_{\rm 2H}^{\rm nonpert.}(k) \equiv P_{\rm 2H}(k) - P_{\rm pert.}(k) .
\label{P-2H-nopert}
\eeq
Of course, the halo model itself is only an approximate and phenomenological
description of the density field. Hence the splitting of nonperturbative contributions
into the two terms $P_{\rm 2H}^{\rm nonpert.}$ and $P_{\rm 1H}$ is not very well and uniquely
defined, especially from a Eulerian point of view. However, from the Lagrangian
point of view that led to the model (\ref{Pk-halos})-(\ref{Pk-2H-1}), the distinction
is easier to make (even though approximate) as pairs of particles belong either to
the same or different halos and we can split their motion into small-scale virial motions
and large-scale collective flows.
Then, the term $P_{\rm 2H}^{\rm nonpert.}$ is the contribution of small-scale
multistreaming to the large-scale power spectrum. This transfer of power is due
to the Fourier transform that defines the power spectrum, which mixes scales:
finite-scale motions contribute to all wave numbers in Eq.(\ref{Pk-def}).
Within our framework (\ref{Pk-2H-1}), this arises from the uncorrelated small-scale
virial motions of both particles in their respective halos [the factor
$\lag e^{\ii\vk\cdot\vx} \rag^{\rm vir}_{q}$ in Eq.(\ref{Pk-2H-1})],
the sticking of particles within pancakes (instead of escaping to infinity as in the
Zel'dovich approximation) [the factors $A_1$ and the complex integral over $y$
in Eq.(\ref{Pk-2H-1})], and the removal of particle pairs that belong to the same
halo [the factor $F_{\rm 2H}$ in Eq.(\ref{Pk-2H-1})] as we collect their contribution
in a separate one-halo term.
At a qualitative level, these factors are analogous to the viscous (sticking in pancakes)
and pressure (uncorrelated small-scale virial motions) terms that have been added
in some previous works \cite{Carrasco2012,Pietroni2012}
to the hydrodynamical equations of motion to build perturbative expansions that can
handle some shell-crossing effects.
These effective approaches, which are based on a separation of scales, 
would provide another route to predict the ``cosmic web'' power spectrum or
two-halo term (\ref{Pk-2H-1}) but cannot describe the fully nonlinear scales
associated with the one-halo term (\ref{Pk-1H}).

\begin{figure}
\begin{center}
\epsfxsize=8.5 cm \epsfysize=6.5 cm {\epsfbox{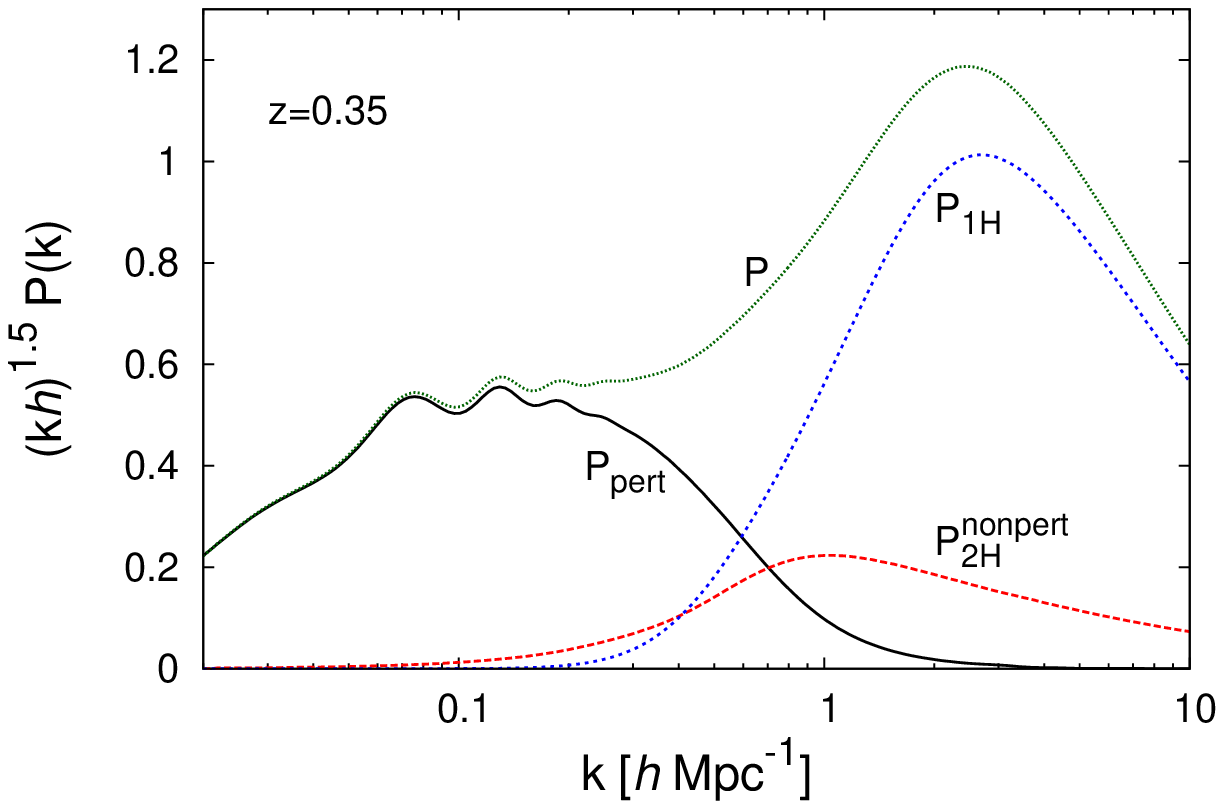}}\\
\epsfxsize=8.5 cm \epsfysize=6.5 cm {\epsfbox{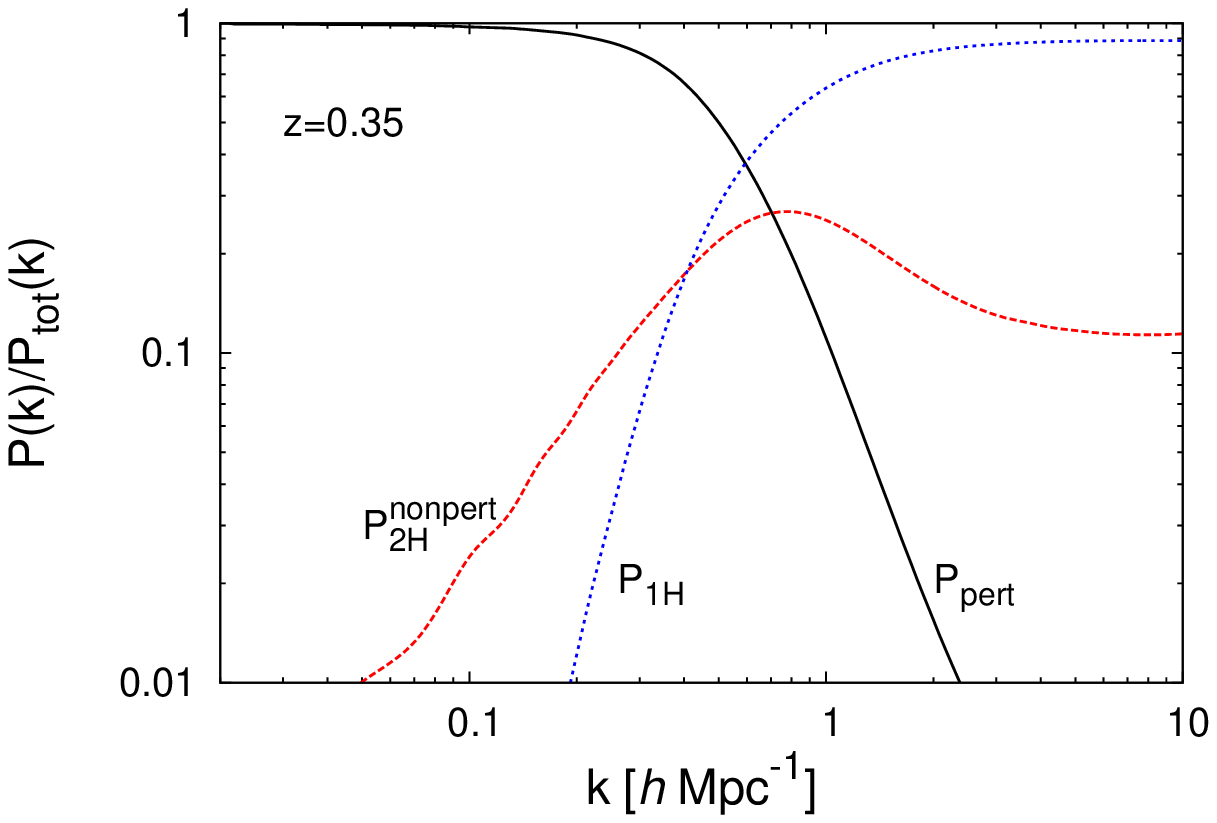}}
\end{center}
\caption{{\it Upper panel:} perturbative and nonperturbative contributions to the
full nonlinear density power spectrum, from Eq.(\ref{P-3split}), at $z=0.35$.
{\it Lower panel:} relative importance of these contributions to the full power
spectrum. $P_{\rm pert.}(k)$ is obtained from Eq.(\ref{P-no-sc-1}), $P_{\rm 1H}(k)$
from Eq.(\ref{Pk-1H}), and $P_{\rm 2H}^{\rm nonpert.}(k)$ from Eqs.(\ref{Pk-2H-1})
and (\ref{P-2H-nopert}).}
\label{fig-kPk_z0.35}
\end{figure}

\begin{figure}
\begin{center}
\epsfxsize=8.5 cm \epsfysize=6.5 cm {\epsfbox{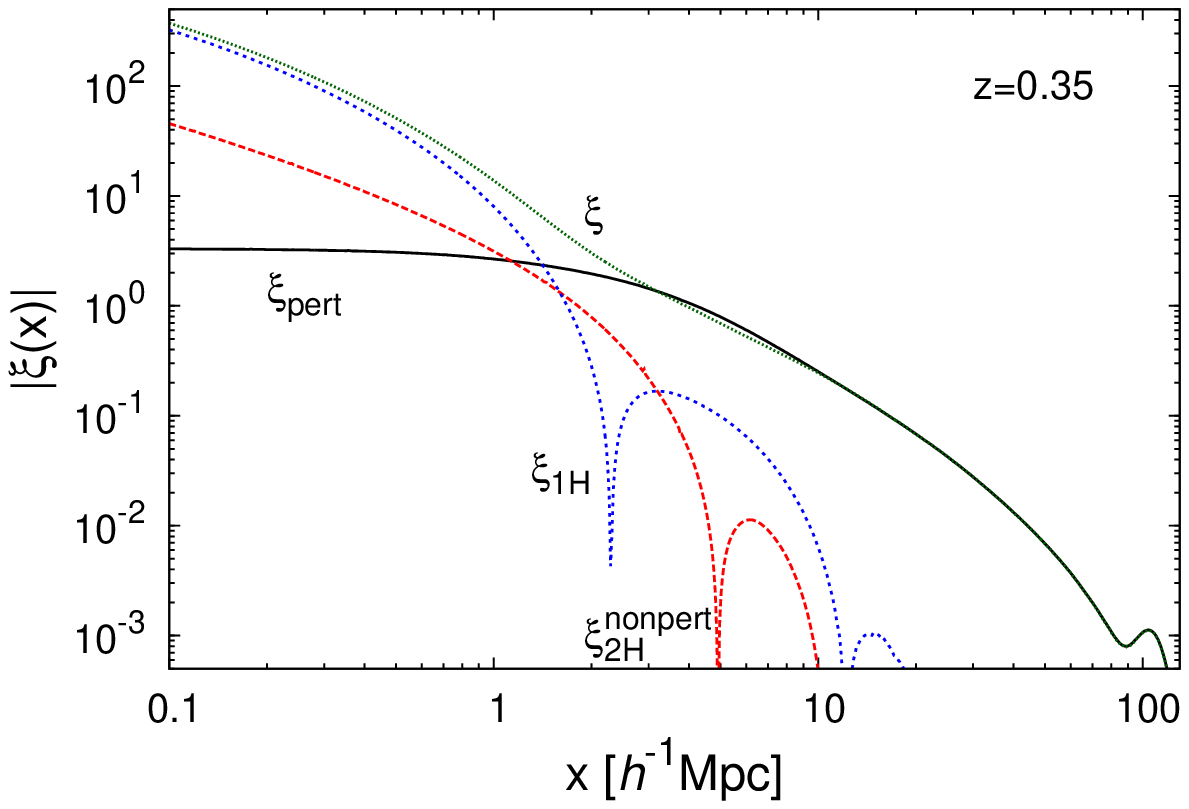}}\\
\epsfxsize=8.5 cm \epsfysize=6.5 cm {\epsfbox{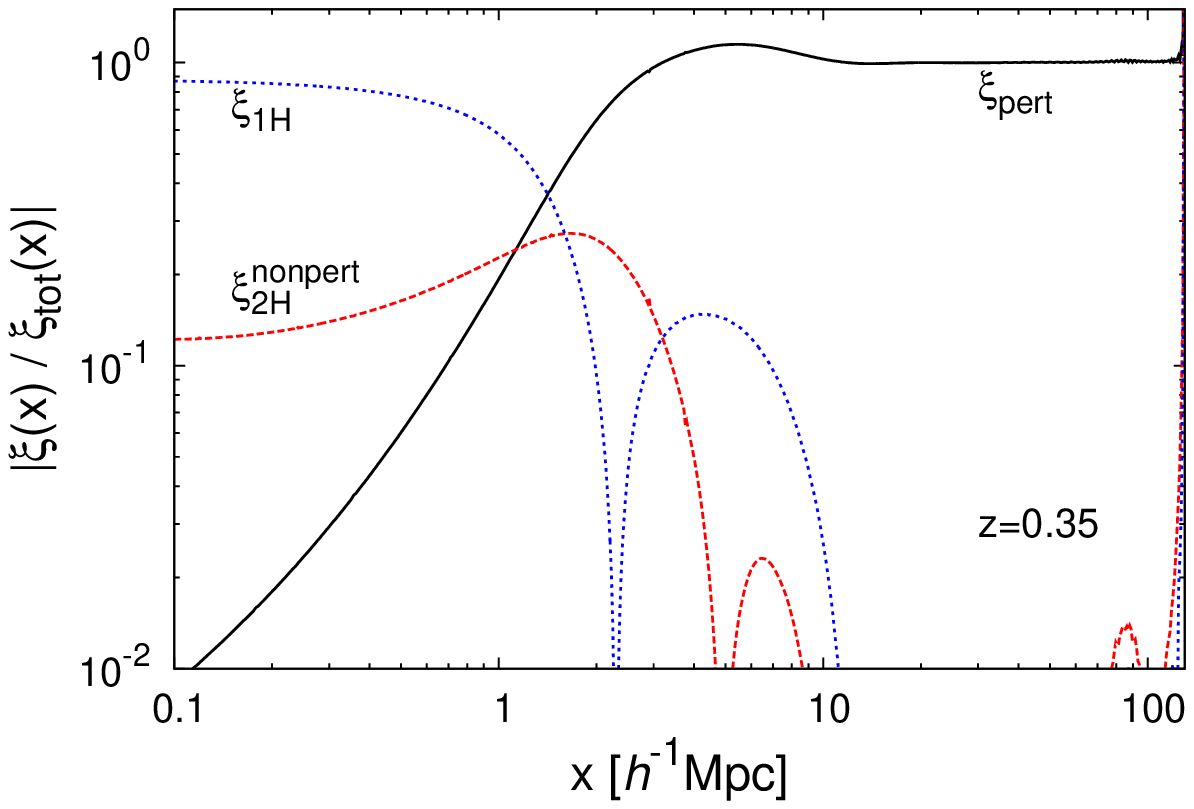}}
\end{center}
\caption{{\it Upper panel:} perturbative and nonperturbative contributions to the
full nonlinear density correlation function, from Eq.(\ref{xi-3split}), at $z=0.35$.
{\it Lower panel:} relative importance of these contributions to the full correlation
function.}
\label{fig-xi_z0.35}
\end{figure}

We show in Fig.~\ref{fig-kPk_z0.35} the three components of Eq.(\ref{P-3split}).
We can check that the perturbative component is dominant on large scales, the
one-halo term on small scales, and the nonperturbative part of the two-halo
term is mainly relevant on intermediate scales.
As seen in the lower panel, the perturbative part $P_{\rm pert.}$ shows a fast decrease
at high $k$ in relative terms. This is because it typically decays faster than $k^{-3}$
as the Zel'dovich power spectrum, because intermediate-scale structures
are erased as particles keep moving on. The one-halo term (\ref{Pk-1H}) shows a fast
decrease at low $k$ because of its $k^4$ tail, ensured by the counterterm $\tW$
in Eq.(\ref{Pk-1H}), associated with mass and momentum conservation.
The term $P_{\rm 2H}^{\rm nonpert.}$, which combines small and large scales, shows
a broader distribution but its precise shape is likely to depend on the details of our model.
In any case, Fig.~\ref{fig-kPk_z0.35} shows that if we require a few percent accuracy
we are sensitive to shell-crossing effects down to $k \sim 0.1 h$Mpc$^{-1}$ at
$z=0.35$.
This agrees within a factor of about 2 with the results obtained in previous
works that used simpler models \cite{Afshordi2007,Valageas2011a}.
As expected, we first encounter the term $P_{\rm 2H}^{\rm nonpert.}$, associated with the
residual effect of small-scale multistreaming onto the large-scale power, and next the
one-halo contribution.

We must note that Fig.~\ref{fig-kPk_z0.35} does not define by itself the limitation
of semianalytical models. This only gives a lower bound to the range of wave numbers
that can be described by semianalytical models, if we set all nonperturbative
contributions to zero and use a Lagrangian-based regularization of perturbation
theory. In practice, if we take into account nonperturbative contributions in an
approximate fashion, as in Ref.\cite{Valageas2013} for instance or by including
some additional pressure terms to the equations of motion, we can extend the
range of validity of semianalytical models. In particular, they only need to be modeled
up to $10\%$ on scales where they do not contribute to more than $20\%$ if we
require a $2\%$ accuracy. 
We discuss in more details these points in Secs.~\ref{mass-concentration} and
\ref{mass-function} below.
However, Fig.~\ref{fig-kPk_z0.35} is useful as a warning to the limitations of
perturbation theories and gives an estimate of the scale and accuracy where
adding high-order contributions to the single-stream perturbative expansions is
relevant.

The Fourier transform of Eq.(\ref{P-3split}) gives the decomposition of the
two-point correlation function,
\beq
\xi(x) = \xi_{\rm pert.}(x) + \xi_{\rm 2H}^{\rm nonpert.}(x) + \xi_{\rm 1H}(x) .
\label{xi-3split}
\eeq
We show our results in Fig.~\ref{fig-xi_z0.35}.
Again, the perturbative term dominates on large scales, the one-halo term on small 
scales, and the nonperturbative part of the two-halo term is mainly relevant on
intermediate scales.
The peaks at $x\simeq 130 h^{-1}$Mpc in the lower panel are due to the zero crossing
of the two-point correlation, which makes the ratios diverge.
The lower panel shows that nonperturbative contributions are negligible at $z=0.35$
on scales larger than $10 h^{-1}$Mpc, even when we require an accuracy of $1\%$.
(This also agrees with Ref.~\cite{Carlson2013} who noticed that the Zel'dovich
correlation function is reasonably accurate down to $\sim 10 h^{-1}$Mpc.)
This shows that real-space statistics provide a robust and efficient probe of cosmology
as they offer a clean separation between perturbative and nonperturbative
contributions.
This is important as the perturbative contributions can be computed from first
principles by systematic expansion schemes, whereas nonperturbative contributions
are necessarily more phenomenological and of limited accuracy.
Moreover, as noticed in Sec.~\ref{Convergence}, the perturbative expansions of the
two-point correlation function show a fast convergence on large scales.
These features can be understood from the physics at play, as nonlinearities arise
from small-scale motions (rather than wave interactions) that take place in configuration
space and do not redistribute matter on large scales.
Then, one can expect a separation of scales to be more readily apparent in
configuration space than in Fourier space, where the integral transform spreads the
contributions to all wave numbers.

\begin{figure}
\begin{center}
\epsfxsize=8.5 cm \epsfysize=6.5 cm {\epsfbox{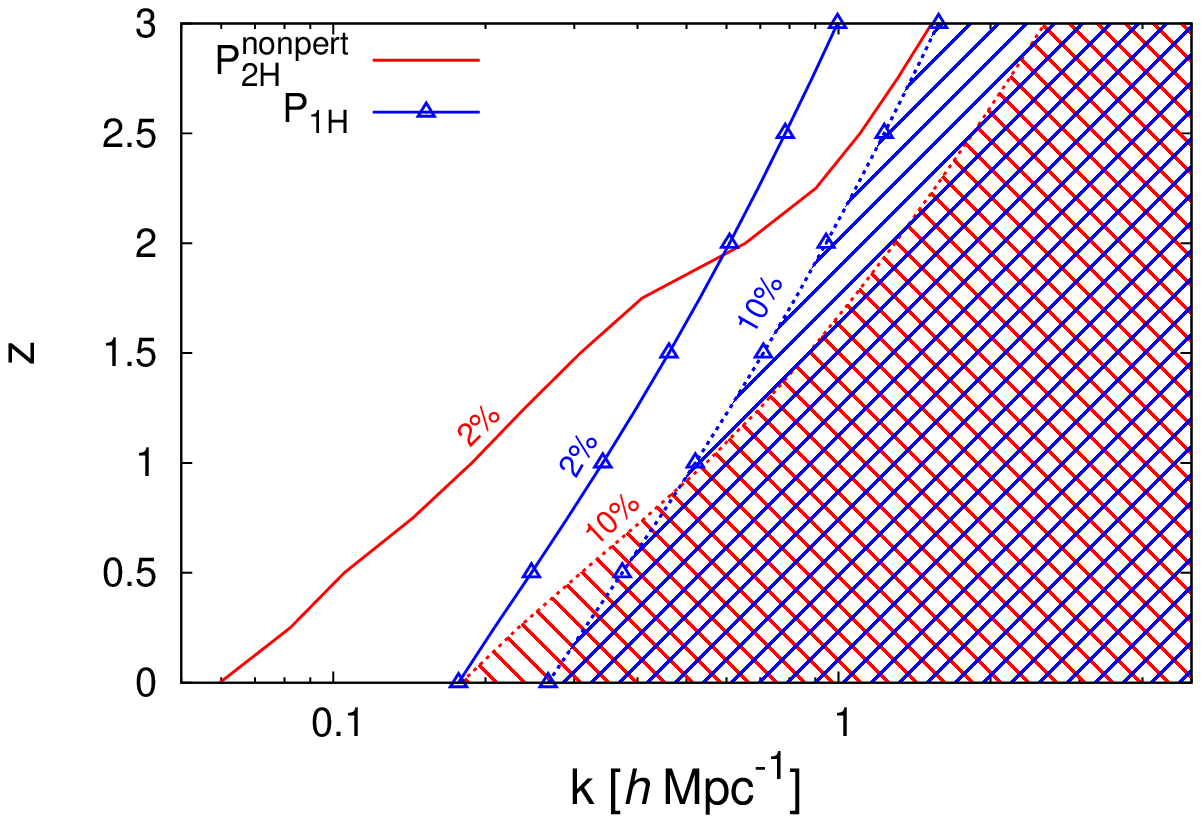}}\\
\epsfxsize=8.5 cm \epsfysize=6.5 cm {\epsfbox{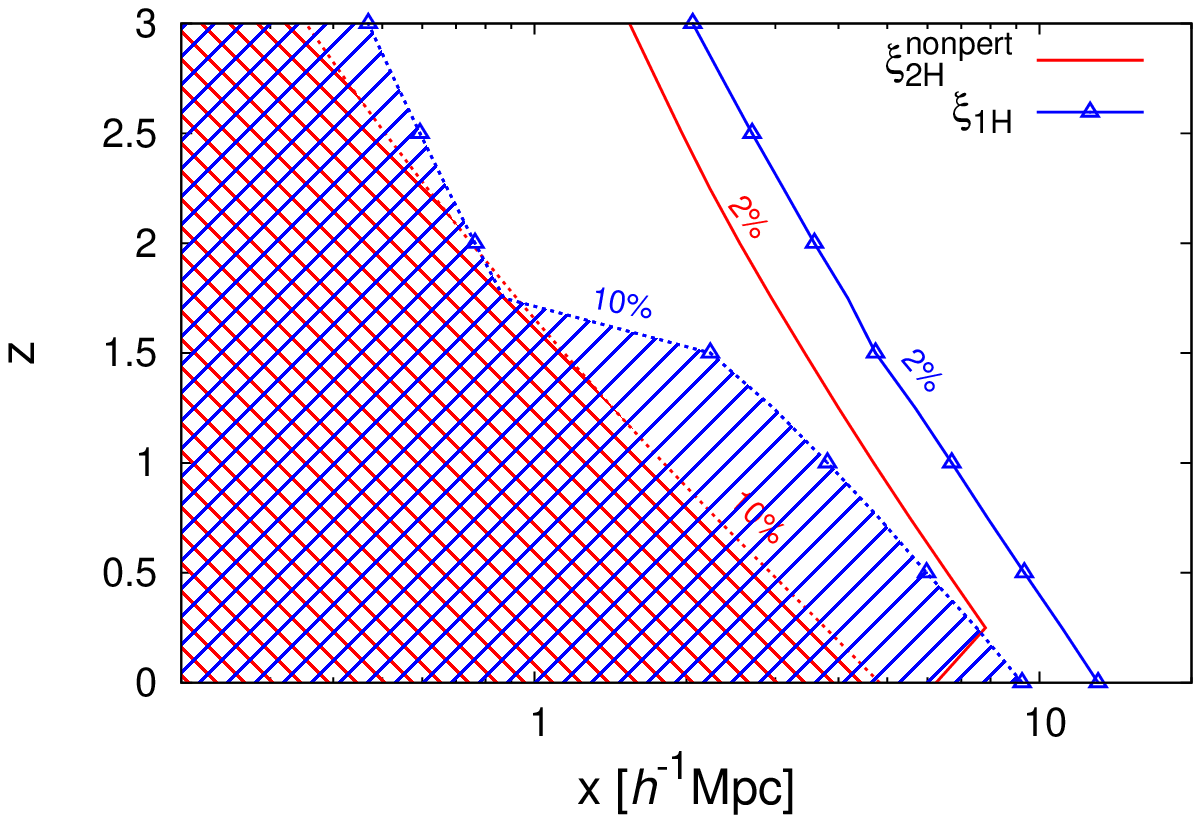}}
\end{center}
\caption{{\it Upper panel:} contour lines in the $(k,z)$-plane of the regions where
$P_{\rm 2H}^{\rm nonpert.}$ or $P_{\rm 1H}$ (lines with triangle symbols) make
more than $2\%$ (solid lines) or $10\%$ (dotted lines) of the full power spectrum.
{\it Lower panel:} similar contour lines in the $(x,z)$-plane for the relative contributions
of $\xi_{\rm 2H}^{\rm nonpert.}$ and $\xi_{\rm 1H}$ to the full two-point correlation $\xi$.}
\label{fig-contours}
\end{figure}

To study the evolution with redshift of these perturbative and nonperturbative
contributions, we show in the upper panel of Fig.~\ref{fig-contours} the contour
lines in the $(k,z)$-plane of the regions where $P_{\rm 2H}^{\rm nonpert.}$ and
$P_{\rm 1H}$ contribute to more than $2\%$ or $10\%$ of the full power spectrum.
The lower panel shows the same contour lines for the two-point correlation in the
$(x,z)$-plane.
At higher redshift the region where the nonperturbative contributions are important
is pushed toward higher wave numbers $k$ and smaller scales $x$.
The contour lines associated with $P_{\rm 2H}^{\rm nonpert.}$ and $P_{\rm 1H}$ are
similar, as they follow the nonlinear scale.
The lower panel shows that if we can model these nonperturbative up to $20\%$
we can reach an accuracy of $2\%$ for $\xi(x)$ down to $x \geq 10$ or
$x \geq 0.5 h^{-1}$Mpc at redshifts $z=0$ or $3$ (provided we have a good
perturbative scheme, but this should not be the limiting factor).
This is rather reassuring, as it ensures a broad range of scales for future wide surveys.

\subsection{Impact of phenomenological parameters}
\label{phenomenological}

As noticed in Sec.~\ref{Impact-of-nonperturbative-contributions}, the study of the
relative importance of nonperturbative contributions to the matter power spectrum
or correlation function only gives a very conservative estimate of the scope of
semianalytical models. Indeed, the latter can include such effects in a
phenomenological manner, e.g. through a halo model, or through an implicit
regularization of the perturbative scheme. 
Then, we can estimate the scope of semianalytical models by studying the sensitivity
of their predictions to the value of these phenomenological parameters, which cannot
be computed by systematic analytical approaches.
Within our framework, this corresponds to investigating the sensitivity of our
predictions on the halo-model parameters, associated with the halo mass function
and density profiles. Indeed, because there is little hope that such properties can
be predicted with a high accuracy by analytical models, they must be taken from
numerical simulations. Then, the theoretical predictions become limited by the
accuracy of these simulations, and more generally by these parameters; even more
so when one considers different cosmologies than those where they were measured.
This yields a limitation to the accuracy of semianalytical models but also of predictions
that can be directly obtained from the numerical simulations themselves.

\subsubsection{Impact of the halo mass-concentration relation}
\label{mass-concentration}

\begin{figure}
\begin{center}
\epsfxsize=8.5 cm \epsfysize=6.5 cm {\epsfbox{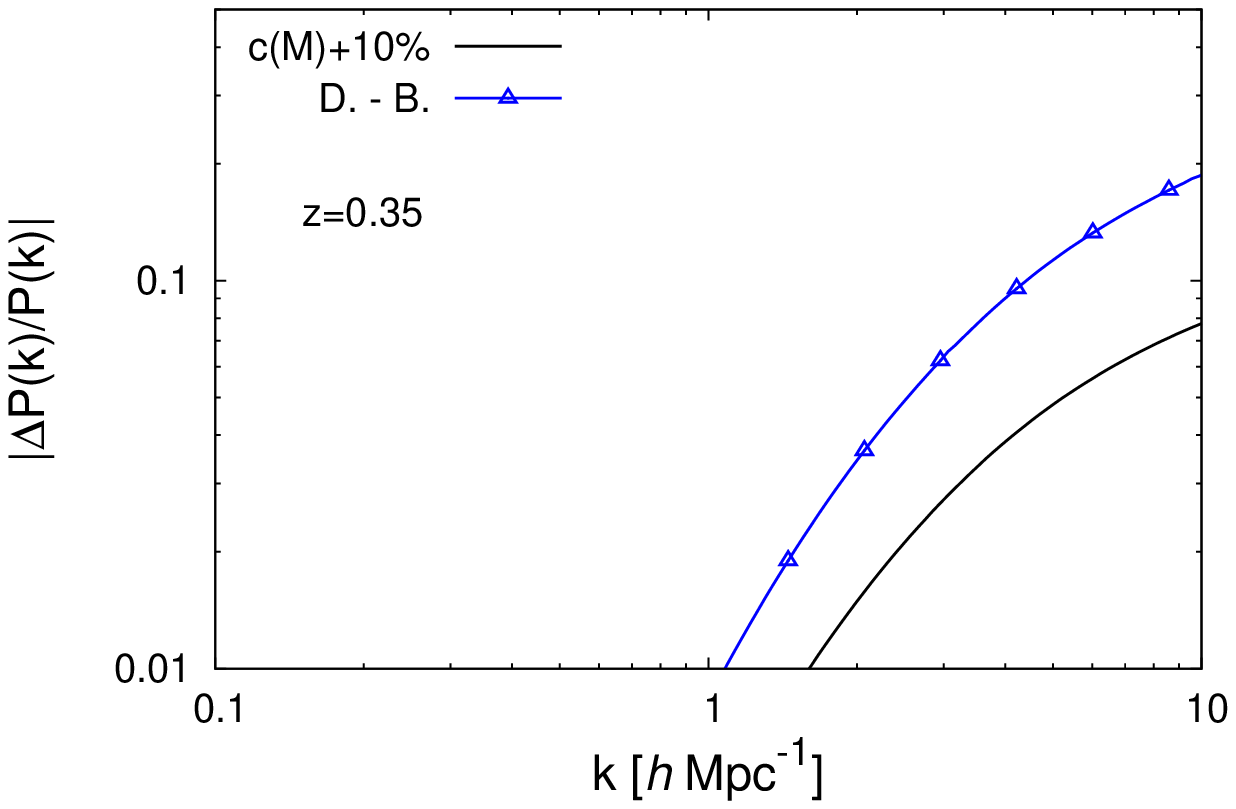}}\\
\epsfxsize=8.5 cm \epsfysize=6.5 cm {\epsfbox{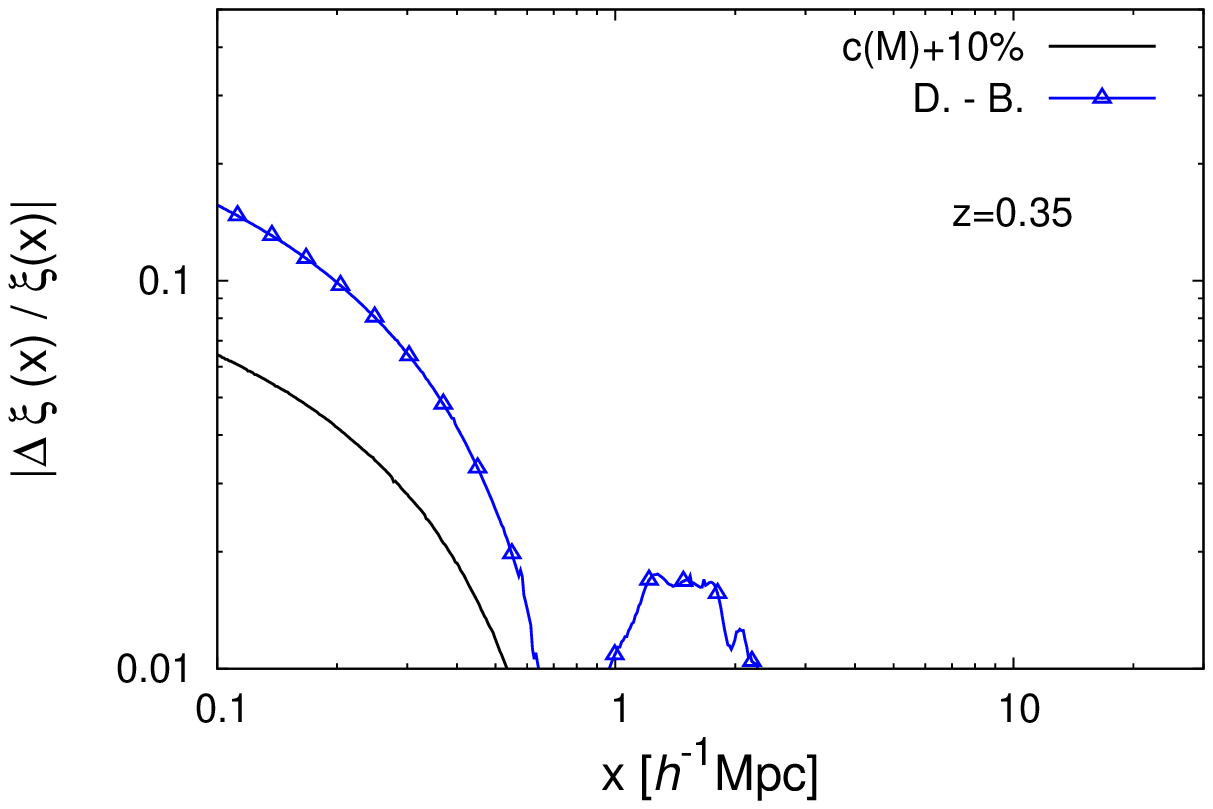}}
\end{center}
\caption{{\it Upper panel:} relative change of the power spectrum at $z=0.35$
when the concentration parameter $c(M)$ is increased by $10\%$,
or when we change from the fit given in Ref.\cite{Duffy2008} to the one of
Ref.\cite{Bhattacharya2013} (label ``D.-B.'' with triangle symbols).
{\it Lower panel:} relative change of the correlation function at $z=0.35$ for the
same cases.}
\label{fig-d_cM_z0.35}
\end{figure}

\begin{figure}
\begin{center}
\epsfxsize=8.5 cm \epsfysize=6.5 cm {\epsfbox{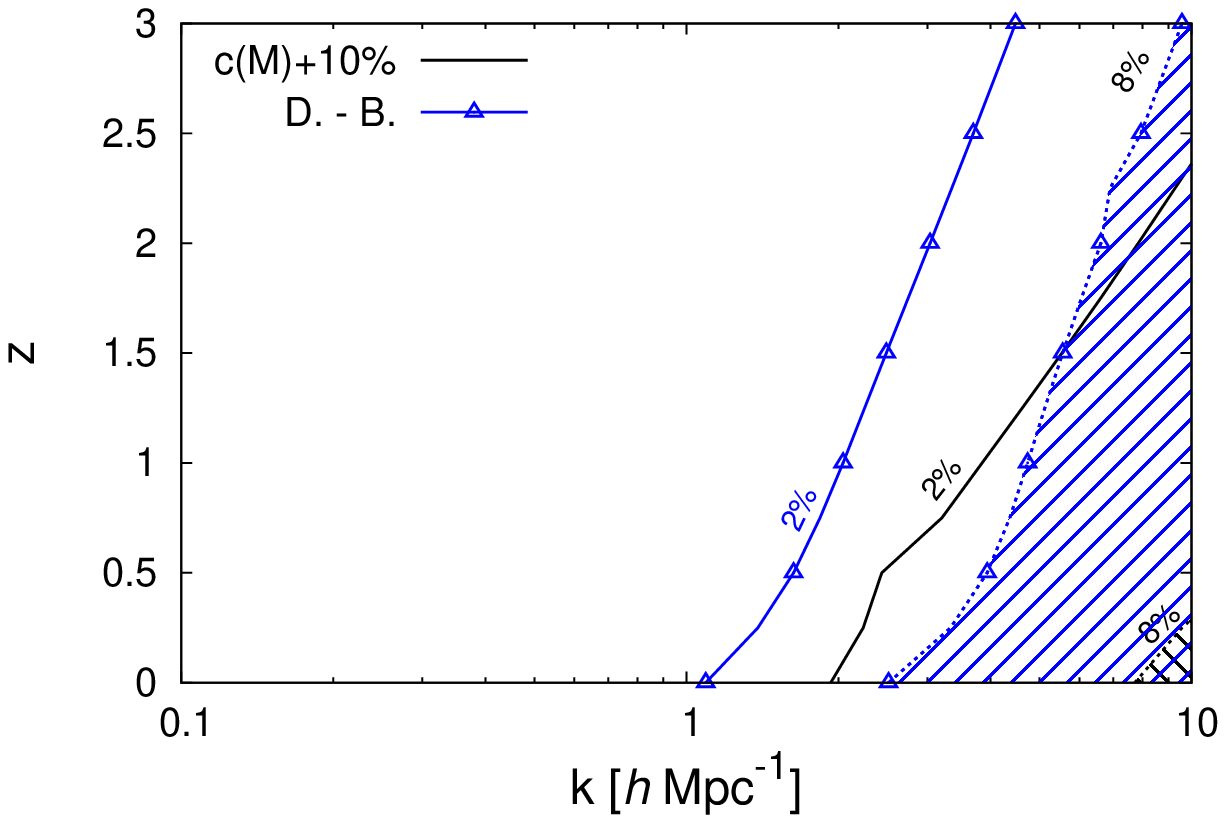}}\\
\epsfxsize=8.5 cm \epsfysize=6.5 cm {\epsfbox{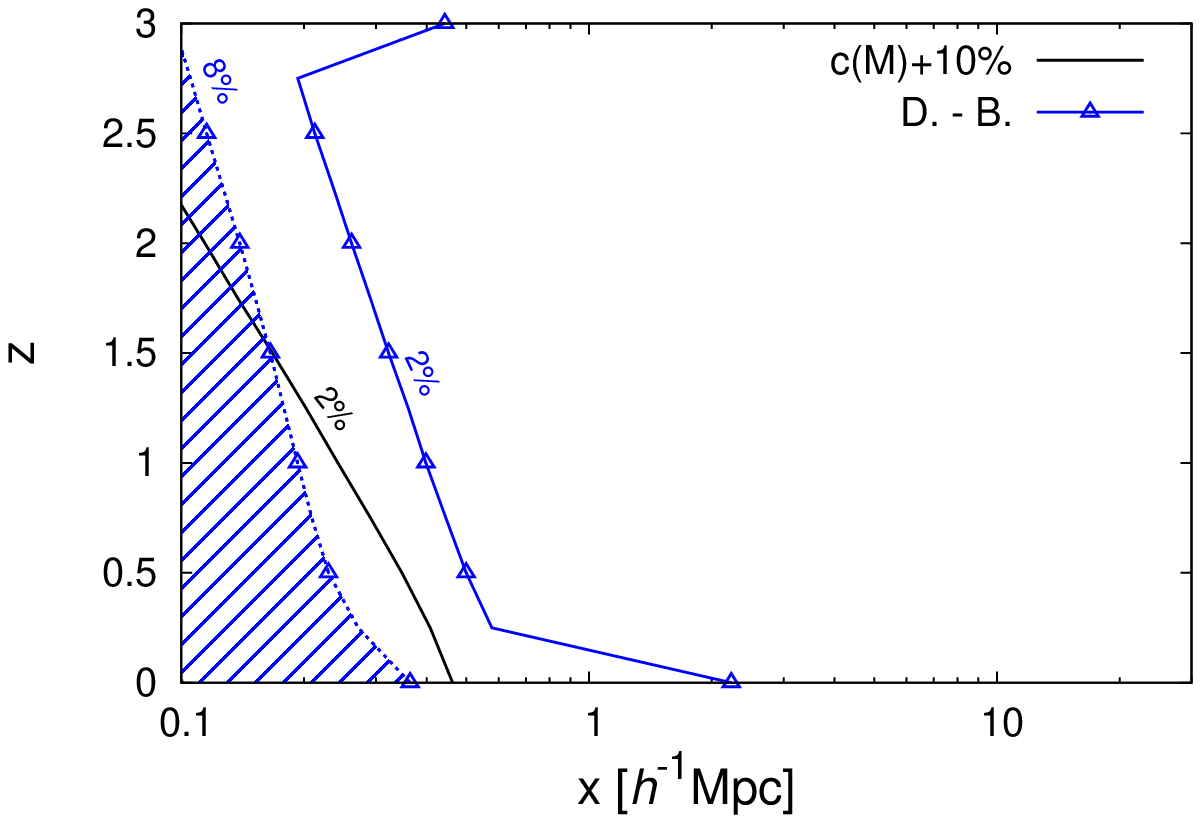}}
\end{center}
\caption{{\it Upper panel:} contour lines in the $(k,z)$-plane of the regions where
the power spectrum is modified by more than $2\%$ (solid lines), or $8\%$
(dotted lines), by a $10\%$ increase of $c(M)$ or by changing from
Ref.\cite{Duffy2008} to Ref.\cite{Bhattacharya2013} (label ``D.-B.'' with triangle
symbols).
{\it Lower panel:} similar contour lines in the $(x,z)$-plane for the
two-point correlation.}
\label{fig-contours_cM}
\end{figure}

The halo density profiles are a first limitation to the accuracy of semianalytical models.
Indeed, being the result of highly nonlinear and nonperturbative processes,
it has proved difficult to obtain accurate and systematic theoretical predictions
(especially within the virial radius).  In practice one uses a typical fitting profile 
(e.g., a rational function) with a few parameters that may depend on the halo mass.
Here, as in \cite{Valageas2013}, we use the NFW profile \cite{Navarro1997}.
The exponents are fixed and there is a single concentration parameter, $c(M)$,
that determines the transition scale between the inner and outer regimes
$\rho \propto x^{-1}$ and $\rho\propto x^{-3}$. 

We show in Fig.~\ref{fig-d_cM_z0.35} the impact on the power spectrum and
correlation function, at $z=0.35$, of a $10\%$ increase of $c(M)$. We also show
the difference between the predictions obtained using two different fits to numerical
simulations from previous works, Refs.\cite{Duffy2008} and \cite{Bhattacharya2013}.
We can see that the difference between published fits for $c(M)$ is of order $10\%$
(somewhat greater; this also depends on mass and redshift). However, it appears
that the impact on the power spectrum and correlation function is restricted to
rather small scales, $k\gtrsim 1 h$Mpc$^{-1}$ and $x\lesssim 2 h^{-1}$Mpc at
$z=0.35$, for a $1\%$ accuracy.

The contour lines in the $(k,z)$ and $(x,z)$ planes of a $2\%$ or $8\%$ impact on
the power spectrum and correlation function are shown in Fig.~\ref{fig-contours_cM},
for these same modifications to $c(M)$.
Again, we find that for the redshift range $0 \leq z \leq 3$ the uncertainties of the
mass-concentration relation only affect the power spectrum and correlation
function on rather small scales.
The comparison with Fig.~\ref{fig-contours} shows that this occurs in the highly
nonlinear regime where the perturbative expansions are no longer valid and
the power spectrum or correlation function is dominated by the one-halo contribution.

Thus, the precise shape of halo profiles should not be a worrying limitation
of semianalytical models, because there remains a large range of scales where its
impact is negligible.

\subsubsection{Impact of the halo mass function}
\label{mass-function}

\begin{figure}
\begin{center}
\epsfxsize=8.5 cm \epsfysize=6.5 cm {\epsfbox{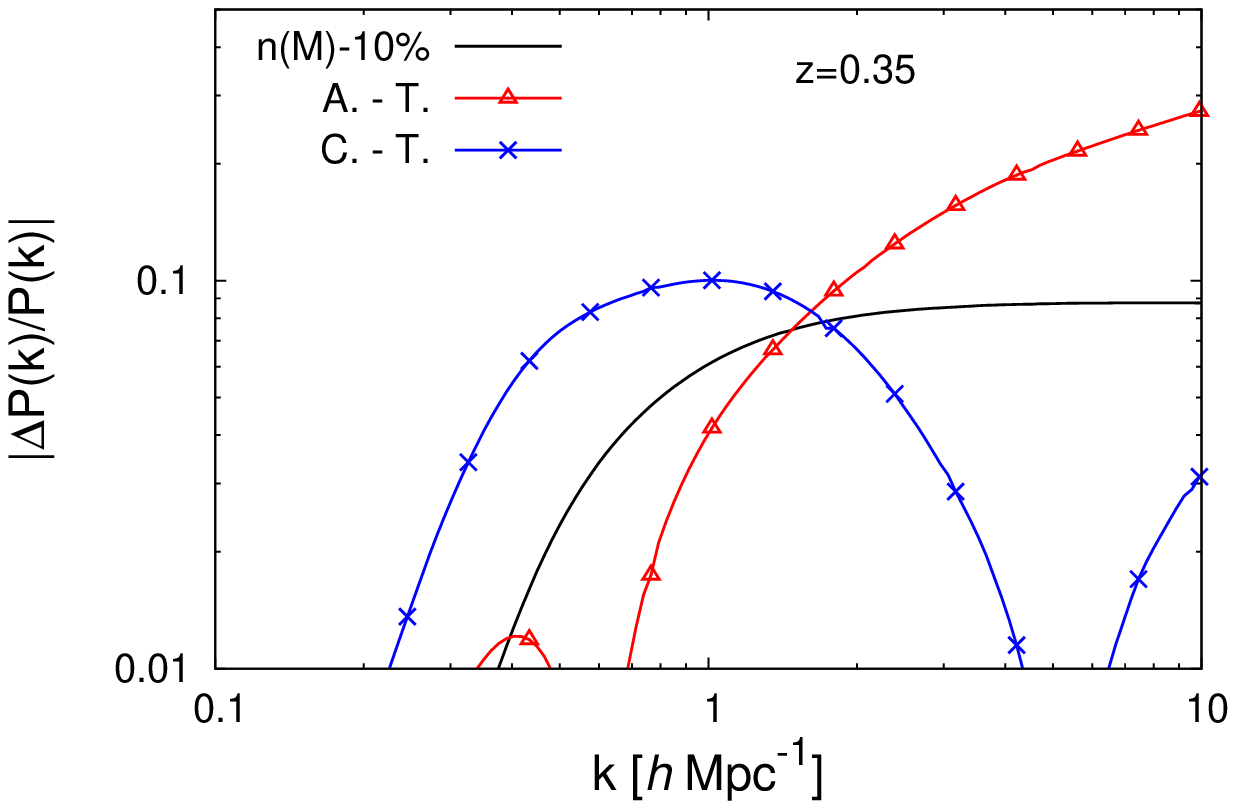}}\\
\epsfxsize=8.5 cm \epsfysize=6.5 cm {\epsfbox{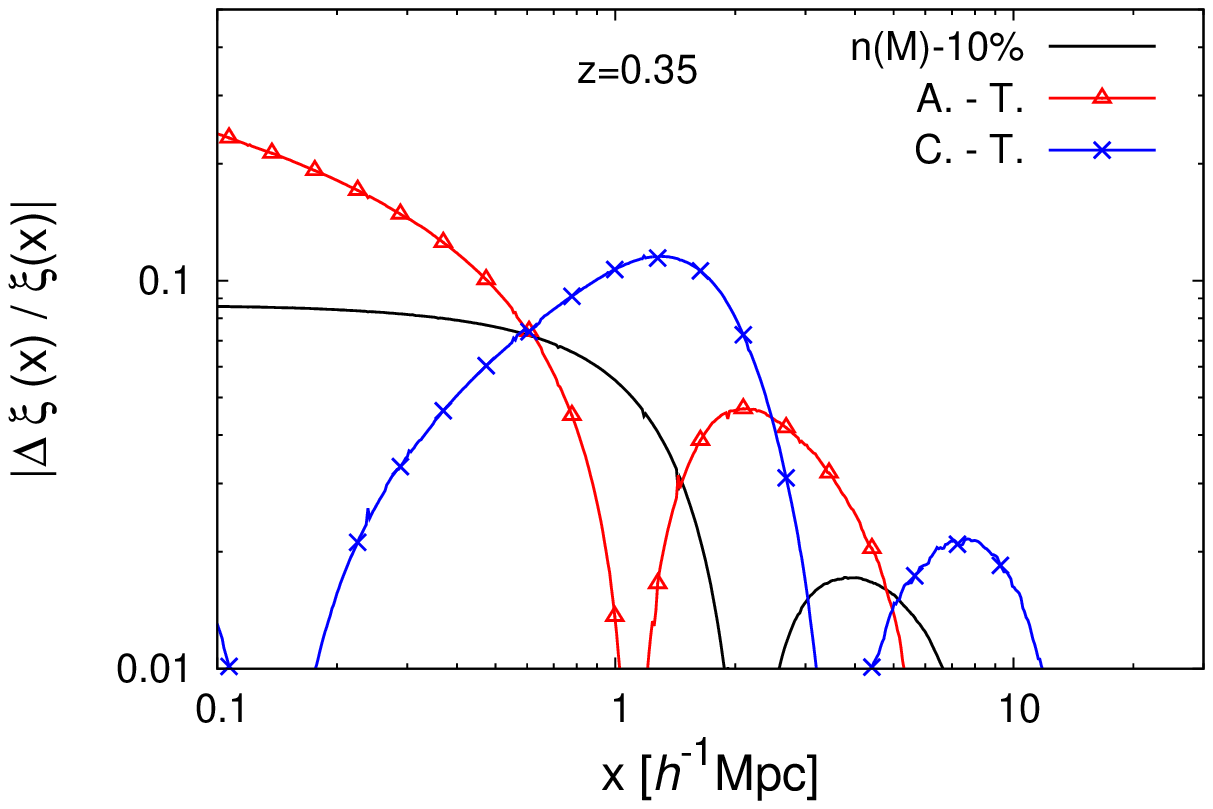}}
\end{center}
\caption{{\it Upper panel:} relative change of the power spectrum at $z=0.35$
when the halo mass function $n(M)$ is decreased by $10\%$,
or when we change from the fit given in Ref.\cite{Tinker2008} to those of
Ref.\cite{Angulo2012} (label ``A.-T.'' with triangle symbols) or Ref.\cite{Courtin2011}
(label ``C.-T.'' with cross symbols).
{\it Lower panel:} relative change of the correlation function at $z=0.35$ for the
same cases.}
\label{fig-d_fM_z0.35}
\end{figure}

\begin{figure}
\begin{center}
\epsfxsize=8.5 cm \epsfysize=6.5 cm {\epsfbox{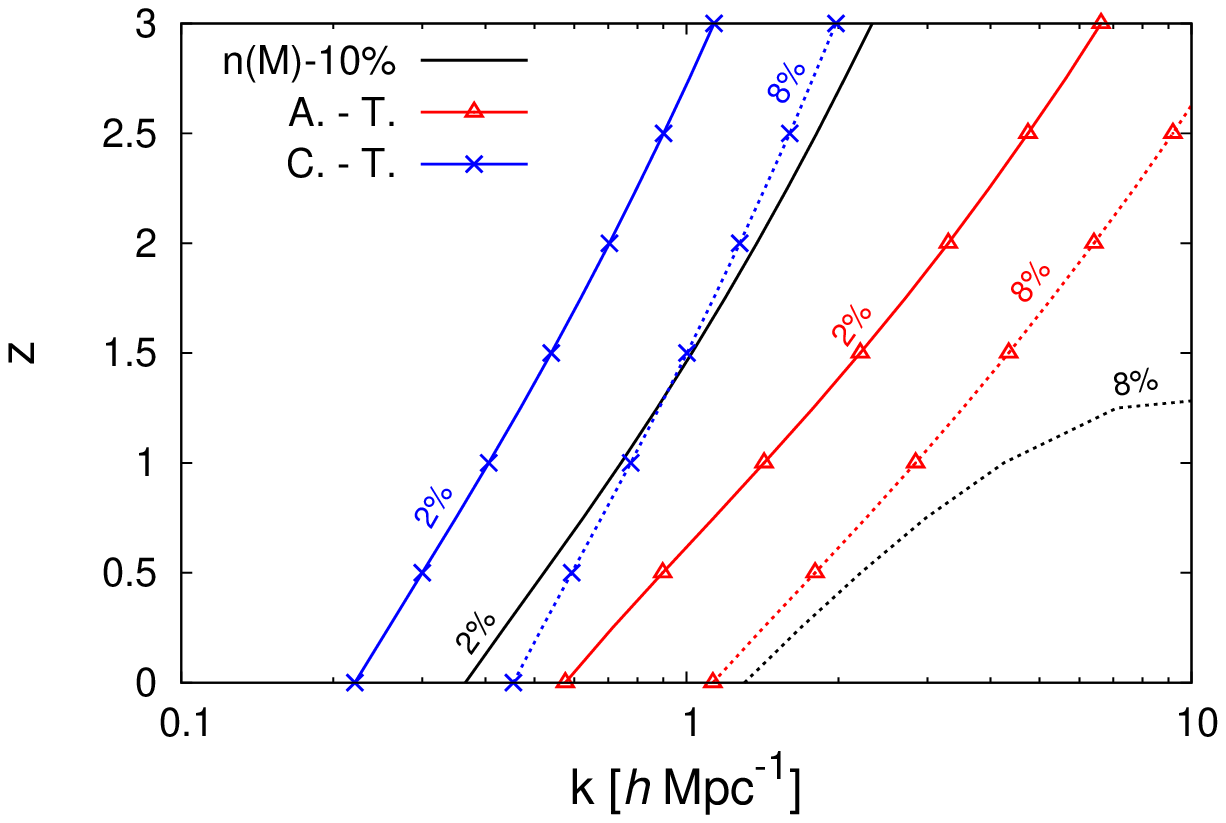}}\\
\epsfxsize=8.5 cm \epsfysize=6.5 cm {\epsfbox{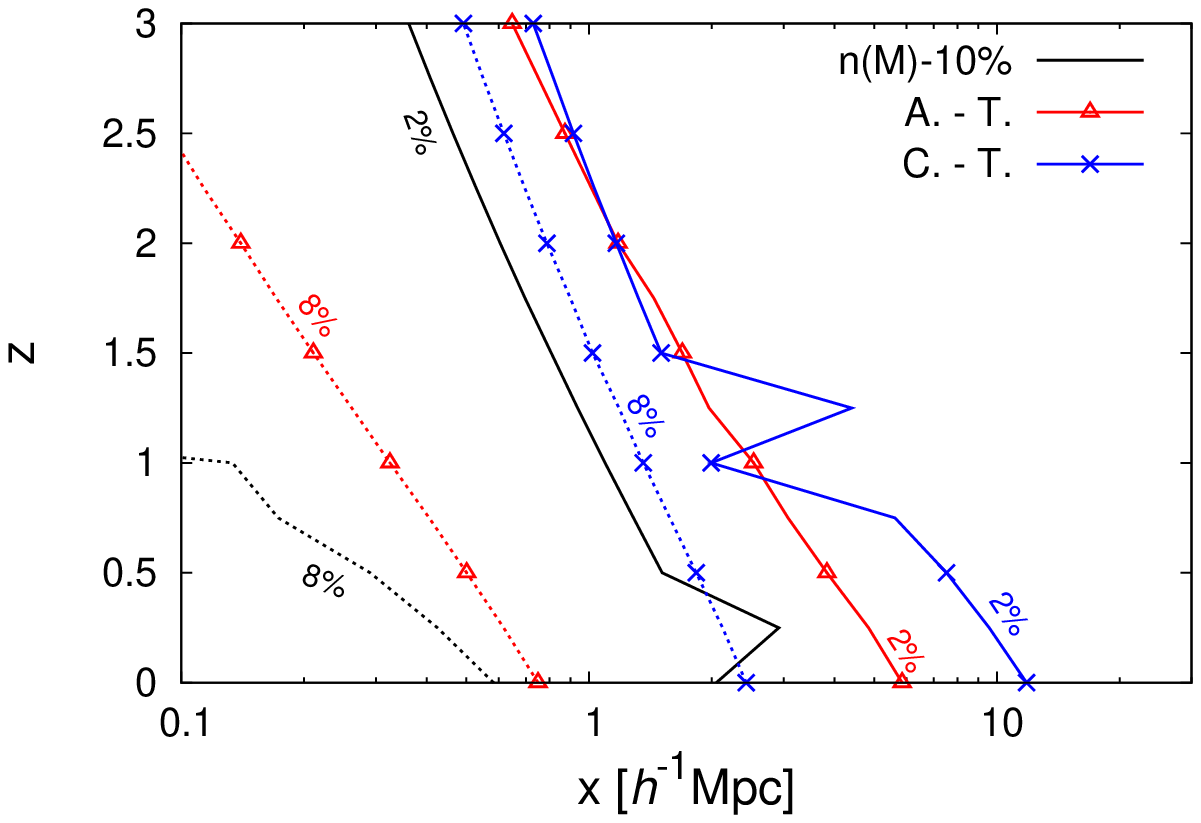}}
\end{center}
\caption{{\it Upper panel:} contour lines in the $(k,z)$-plane of the regions where
the power spectrum is modified by more than $2\%$ (solid lines), or $8\%$
(dotted lines), by a $10\%$ decrease of $n(M)$, or by changing from 
Ref.\cite{Tinker2008} to Ref.\cite{Angulo2012} (label ``A.-T.'' with triangle symbols)
or to Ref.\cite{Courtin2011} (label ``C.-T.'' with cross symbols).
{\it Lower panel:} similar contour lines in the $(x,z)$-plane for the
two-point correlation.}
\label{fig-contours_fM}
\end{figure}

Apart from the halo profiles, a second limitation to the accuracy of semianalytical 
models is the halo mass function itself. In principle, it should be more easily predicted 
than halo profiles, because one does not need to follow the late virialization stages
of inner halo regions but only to count collapsed regions. This explains the
relative success of various analytical approaches \cite{Press1974,Paranjape2013}
that try to detect future halos from the initial linear density field (for the high-mass tail).
However, it has proved difficult to go below a $20\%$ accuracy (this depends on
mass and redshift) and most works use fits to numerical simulations, or involve
some parameters that are taken from simulations.

We show in Fig.~\ref{fig-d_fM_z0.35} the impact on the power spectrum and
correlation function, at $z=0.35$, of a $10\%$ decrease of the mass function $n(M)$.
We also show the difference between the predictions obtained using three different fits
to numerical simulations, Refs.\cite{Tinker2008}, \cite{Angulo2012}, 
and \cite{Courtin2011}.
We can see that the difference between published fits for $n(M)$ is of order $10\%$
(somewhat greater; this also depends on mass and redshift).
The comparison with Fig.~\ref{fig-d_cM_z0.35} shows that the impact of a
$10\%$ inaccuracy of the halo mass function is greater than the impact of a
$10\%$ inaccuracy of the mass-concentration relation.
Therefore, this could be the limiting factor of semianalytical models.

Decreasing (or increasing) the halo mass function by $10\%$ is not realistic
because the total halo mass fraction should remain at unity (or at least not greater
than unity). However, the power spectrum and correlation functions on large scales
are mostly sensitive to massive and large halos, so that the constraint of a
unit normalization (which is satisfied by the three fits from
Refs.\cite{Tinker2008,Angulo2012,Courtin2011}) is not sufficient to lessen the impact
on the power spectrum.
The large-mass tail is also difficult to measure from numerical simulations, because
these are rare objects.
These fits for halo mass functions are actually defined in different manners, as
one can use different halo-finder algorithms (e.g., based on a spherical-overdensity
criterion or friends-of-friends procedures) and different halo definitions (e.g., different
halo density contrasts). This is a further difficulty for semianalytical models, as
different definitions may be relevant for different purposes.

In any case, Fig.~\ref{fig-d_fM_z0.35} shows that there remains a significant range
of scales that is not affected by these inaccuracies of the halo mass function.
In particular, for the correlation function at $z=0.35$ scales beyond $10 h^{-1}$Mpc
are not affected at the percent level.
This means that accurate theoretical predictions can be obtained for a useful
range of scales and redshifts.

The contour lines in the $(k,z)$ and $(x,z)$ planes of a $2\%$ or $8\%$ impact on
the power spectrum and correlation function are shown in Fig.~\ref{fig-contours_fM},
for these same modifications to $n(M)$.
We can check from the comparison with Fig.~\ref{fig-contours} that these effects
only occur on small scales where the nonperturbative contributions are significant.
Again, the comparison with Fig.~\ref{fig-contours_cM} shows that 
in the redshift range $0 \leq z \leq 3$ the uncertainties of the
halo mass function have a greater impact than those of the 
mass-concentration relationship.
The real-space correlation function again appears to be more robust and provides
a cleaner separation from these effects than the power spectrum.

The sources of uncertainty displayed in Figs.~\ref{fig-d_fM_z0.35} and
\ref{fig-contours_fM} also apply to the power spectra and correlation functions
directly measured from numerical simulations. Our analysis describes how
the measures of halo mass functions and power spectra from simulations are correlated.

\section{Conclusion}
\label{Conclusion}

In this paper, using an accurate description of the matter power spectrum that
combines perturbation theory with a halo model, we have investigated
the possible accuracy that can be expected from semianalytical models.

First, focusing on the perturbative component, we have found that the simple
reorganization of the standard perturbation theory with a Gaussian damping
prefactor provides a well-ordered convergence for the power spectrum at
low $k$. It also provides a finite two-point correlation function that is accurate
at the percent level on BAO scales as soon as we go up to order $P_L^2$.
Lagrangian-space expansions appear more efficient than their Eulerian
counterparts when both are truncated at a low order, $N \leq 4$, but at high
orders the convergence is no longer well ordered and shows signs of divergence
at high $k$. On the other hand, the correlation function obtained from these
Lagrangian-space expansions is also accurate at the percent level on BAO scales
as soon as we go up to order $P_L^2$ (and $N \lesssim 4$).

Second, we have investigated the importance of nonperturbative contributions
to the power spectrum. Those coming from the two-halo term, which may be seen
as a backreaction of small scales onto large scales, affect a rather large range of
wave numbers if one looks for a percent accuracy for $P(k)$. Those coming
from the one-halo term, which correspond to inner halo regions, are restricted to
higher wave numbers although they already reach a percent level at
$k \sim 2 h$Mpc$^{-1}$ at $z=0.35$. The separation between perturbative and
nonperturbative effects appears to be better defined in configuration space.
Thus, all nonperturbative effects are smaller than $1\%$ at all redshifts
on scales $x \gtrsim 10 h^{-1}$Mpc, for the correlation function.
This may be understood from the fact that
these nonperturbative processes (shell crossing and virialization within halos)
occur through local processes in real space, rather than wave interactions in Fourier
space.

These estimates of the scales where nonperturbative effects are non-negligible
can also be useful when one compares perturbative schemes with numerical
simulations, to avoid meaningless comparisons.
Indeed, whereas most perturbative schemes do not include shell-crossing
effects, numerical simulations include all contributions at once and a
good agreement on scales where the latter are not negligible can be misleading. 

The relative importance of such nonperturbative effects is not necessarily a
limit to semianalytical models if they can be accurately described. To assess
the actual accuracy of semianalytical modeling, we have then estimated the
impact on the power spectrum of the uncertainty of the mass-concentration
relation and of the halo mass function. These uncertainties apply as well to
the predictions obtained from the numerical simulations themselves.
We find that the current accuracy
of the mass-concentration relation (of order $10\%$) is not a worrying limitation
to theoretical predictions, as it only yields an uncertainty below the percent level
on large scales with $k \lesssim 1 h$Mpc$^{-1}$ or $x \gtrsim 1  h^{-1}$Mpc
at all redshifts. This is due to the fact that changes to halo profiles only
make a small-scale redistribution of matter and do not modify large-scale properties.
The current uncertainty on the halo mass function is a greater problem for
semianalytical models as it can affect wave numbers down to
$k \sim 0.2 h$Mpc$^{-1}$ and scales up to $x \sim 10 h^{-1}$Mpc at $z=0$
if we require a percent accuracy. Indeed, this corresponds to a reorganization of
matter on the scale of the largest halos, because weakly nonlinear scales are mostly
sensitive to the largest halos and the constraint associated with the normalization
to unity of the halo mass function is not sufficient to damp this effect.
Again, it appears that configuration-space statistics are better suited to separate
such effects. In particular, while $k \sim 0.2 h$Mpc$^{-1}$ is close to the scales
($k \sim 0.1 h$Mpc$^{-1}$) where baryon acoustic oscillations can be measured
in the power spectrum, $x \sim 10 h^{-1}$Mpc is quite far from the
correlation-function acoustic peak ($x \sim 105 h^{-1}$Mpc).

These results also apply to the power spectra measured in numerical simulations,
as they describe how uncertainties of the halo mass functions and power spectra
measured in these simulations are correlated.

From the observational point of view, one must then balance the higher
accuracy of the theoretical predictions in configuration space with the easier
handling of Fourier-space data, because of their better-behaved covariance
matrices (which are diagonal in the linear regime because different wave numbers
are uncorrelated).

\begin{acknowledgments}

This work is supported in part by the French Agence Nationale de la Recherche under Grant ANR-12-BS05-0002.

\end{acknowledgments}

\appendix

\section{Appendix: Combining the halo model with one-loop perturbation theory}

In this appendix we provide some more details about the combination of one-loop
perturbation theory and halo model that defines the nonlinear
power spectrum that we use in this paper, given by
Eqs.(\ref{Pk-halos})-(\ref{Pk-2H-1}). See Ref.\cite{Valageas2013} for details.

In the Lagrangian-space framework, particles follow trajectories
$\vx_i(t)=\vq_i+ \Psi_i (t)$, where $\vq_i$ is the initial position and $\Psi_i$
the displacement field. At linear order, the variances of the relative displacement
$\Psi=\Psi_2-\Psi_1$ of two particles $1$ and $2$, in the transverse and longitudinal
directions with respect to the initial separation vector $\vq=\vq_2-\vq_1$, are
\beq
\sigma^2_{\parallel}(q) = \lag \Psi_{L\parallel}^2 \rag = 2 \int \dd\vk \; 
[1-\cos (k_\parallel q)] \frac{k^2_\parallel}{k^4} P_L (k) ,
\label{sig-parallel}
\eeq
\beq
\sigma^2_{\perp}(q) =  \lag \Psi_{L\perp}^2 \rag = 2 \int \dd\vk \; 
[1-\cos (k_\parallel q)] \frac{k^2_\perp}{k^4} P_L (k) ,
\label{sig-perp}
\eeq
where $k_{\perp}$ is the component along one of the two transverse directions.
The power spectrum in the Zel'dovich approximation is obtained by considering the
displacement field at linear order.
For Gaussian initial conditions, this yields a Gaussian distribution for the relative
displacements and, using the exact expression (\ref{Pk-cum}), this leads to the
Zel'dovich power spectrum (\ref{PZ-def}),
where $\mu= (\vk\cdot\vq)/(k q)$.
We go beyond the Zel'dovich approximation by including nonlinearities in the 
distribution of the parallel displacement field, while keeping linear theory for the
transverse one.
Thus, introducing the rescaled longitudinal relative displacement $\kappa$ of the
pair of particles, and its linear variance $\sigma^2_{\kappa}$,
\beq
\kappa = \frac{x_{\parallel}}{q} , \;\;\;
\sigma^2_{\kappa} = \frac{\sigma_{\parallel}^2}{q^2} ,
\eeq
we define its cumulant generating function $\varphi(y)$ by
\beq
\lag e^{-y \kappa/\sigma^2} \rag = e^{-\varphi(y)/\sigma^2_{\kappa}} .
\label{phi-par-def}
\eeq
By definition, this generating function must satisfy the series expansion
(\ref{phi-cum}), with the first few orders given by Eq.(\ref{phi-y-0}).
Then, we use the ansatz
\beq
\varphi(y) = \frac{1-\alpha}{\alpha} 
\left(1+\frac{y}{1-\alpha}\right)^{\alpha} - \frac{1-\alpha}{\alpha} ,
\label{phi-alpha-def}
\eeq
where the scale-dependent parameter $\alpha(q)$ is given by
\beq
\alpha(q) = \frac{2-S_3^{\kappa}(q)}{1-S_3^{\kappa}(q)} .
\label{S3-alpha}
\eeq
This is the simplest function that ensures consistency with the constraint (\ref{phi-y-0})
[and that $-\varphi(y)$ be convex, which must be satisfied to
provide a meaningful cumulant generating function].
The associated power spectrum $P_{\rm pert.}(k)$ is given by Eq.(\ref{P-no-sc-1}),
using Eq.(\ref{phi-par-def}) into Eq.(\ref{Pk-def}) and recalling that we keep linear
theory for the transverse displacement.
Moreover, $P_{\rm pert.}(k)$ is exact up to one-loop order (i.e., up to order
$P_L^2$) provided $S_3^{\kappa}(q)$ is given by
\beqa
S_3^{\kappa}(q) &=& -\frac{24\pi}{\sigma_{\kappa}^4} \int_0^{\infty} \dd k \;
\frac{P^{\rm 1 loop}(k)- P_{\rm Z}^{\rm 1 loop}(k)}{q^4 k^2}\nonumber\\
&& \times \left[ 2+ \cos(kq) - 3 \frac{\sin (kq)}{kq} \right] ,
\label{S3-def}
\eeqa
where $P^{\rm 1 loop}(k)$ is the exact one-loop power spectrum constructed with
standard perturbation theory, whereas $P_{\rm Z}^{\rm 1 loop}(k)$ is the one-loop
power spectrum obtained from the Zel'dovich power spectrum (\ref{PZ-def}).
Thus, the power spectrum (\ref{P-no-sc-1}) is a generalization of the Zel'dovich power
spectrum.
It is consistent with the exact perturbative expansion up to one-loop order
(i.e., $P_L^2$), whereas the Zel'dovich power spectrum only agrees at linear order,
and it also contains some perturbative terms at all higher orders in both Eulerian and
Lagrangian spaces [generated through the nonpolynomial function $\varphi(y)$ and
the exponential in Eq.(\ref{P-no-sc-1})].

The generating function $\varphi(y)$ of Eq.(\ref{phi-par-def}) also defines the
probability distribution function of $\kappa$,
\beq
{\cal P}_{\varphi}(\kappa)= \int_{-\ii\infty}^{\ii\infty} \frac{\dd y}{2\pi \ii
\sigma^2_{\kappa}} \; e^{[\kappa y -\varphi(y)]/\sigma_{\kappa}^2} .
\label{Pkap-par}
\eeq
The perturbative expression (\ref{P-no-sc-1}) does not take into
account nonperturbative phenomena such as shell crossings, which can be
approximated using a simplified adhesion model whereby particles coalesce when
$\kappa<0$. This is described by modifying the probability distribution as
\beq
{\cal P}^{\rm ad.}(\kappa)= a_1 \, \Theta (\kappa >0)
{\cal P}_{\varphi}(\kappa)+ a_0 \, \delta_D(\kappa) ,
\label{Pkap-ad}
\eeq
where $a_{0,1}$ are determined by the constraints
$\lag 1 \rag= \lag \kappa \rag=1$.
This provides a simplified account of the formation of pancakes (the first
nonperturbative structures on large scales, such as the ``walls'' around cosmic
voids or underdense regions), and it leads to the ``cosmic web'' power spectrum
\beqa
P_{\rm c.w.}(k) \!\!\! & = & \!\!\!\! \int \!\! \frac{\dd\vq}{(2\pi)^3} \;
\frac{1}{1\!+\!A_1} \; e^{-\frac{1}{2} k^2 (1-\mu^2) \sigma_{\perp}^2}  
\biggl \lbrace e^{-\varphi(-\ii k q \mu \sigma^2_{\kappa}) /\sigma_{\kappa}^2}
\nonumber \\
&& \hspace{-1cm} + A_1 + \int_{0^+-\ii\infty}^{0^++\ii\infty} \frac{\dd y}{2\pi\ii} \;
e^{-\varphi(y)/\sigma^2_{\kappa}}
\left( \! \frac{1}{y} - \frac{1}{y\! + \! \ii k q \mu \sigma^2_{\kappa}} \! \right)
\!\! \biggl \rbrace , \nonumber \\
&& \label{Pk-cw}
\eeqa
where $A_1=(1-a_1)/a_1$ is given by
\beq
A_1 = \sigma^2_{\kappa} \int_{0^+-\ii\infty}^{0^++\ii\infty} \frac{\dd y}{2\pi\ii y^2} \;
e^{-\varphi(y)/\sigma^2_{\kappa}} .
\label{A1-def}
\eeq
The power spectra (\ref{P-no-sc-1}) and (\ref{Pk-cw})
are identical to all orders of perturbation theory, and only differ by nonperturbative
corrections of the form $e^{-1/\sigma^2}$ associated with the adhesionlike
modification (\ref{Pkap-ad}).

To go to highly nonlinear scales, we use the halo model and the power spectrum
is split over one-halo and two-halo components as in Eq.(\ref{Pk-halos}).
Then, the probability that two particles of initial separation $q$ belong to the same halo
of mass $M$ is 
\cite{Valageas2011d}
\beq
F_{\rm 1H}(q)= \int_{\nu_{q/2}}^\infty \frac{d\nu}{\nu} f(\nu) 
\frac{(2q_M - q)^2(4q_M +q)}{16 q_M^3} ,
\eeq
where $f(\nu)$ is the scaling function that determines the halo mass function,
$\nu= \delta_L(M)/\sigma_M$, and $M= 4\pi\bar\rho q_M^3 /3$. 
(The lower bound of the integral corresponds to the mass enclosed within a radius
$q/2$.)
The probability of belonging to two halos is $F_{\rm 2H}=1-F_{\rm 1H}$.
Finally, the average of the component of the particle displacements that is associated
with small-scale virialized motions within halos reads as
\beq
\lag e^{\ii \vk\cdot\vx} \rag^{\rm vir}_q= \left[ \frac{\int_0^{\nu_{q/2}} \frac{d\nu}{\nu}
f(\nu) \tilde u_M(k)}{\int_0^{\nu_{q/2}} \frac{d\nu}{\nu} f(\nu)} \right]^2 ,
\label{virial}
\eeq
because we assume that virialized motions within two different halos are uncorrelated.
Here, we have defined the Fourier transform of the halo profile as
\beq
\tilde{u}_M(k)= \frac{\int \dd\vx \; e^{-i\vk\cdot\vx} \rho_M(x)}{\int \dd\vx \; \rho_M(x)} ,
\label{u-M-def}
\eeq
with $M=\int \dd\vx \, \rho_M (x)$.

Then, the two-halo part $P_{\rm 2H}(k)$ of the power spectrum is given by
Eq.(\ref{Pk-2H-1}), where we recognize the ``cosmic web'' power spectrum
(\ref{Pk-cw}), to which we have added the factor $F_{\rm 2H}$, to avoid
double counting with the one-halo term, and the small-scale motions factor
(\ref{virial}), to take into account the finite width of halos.
The one-halo part $P_{\rm 1H}(k)$ is given as usual by Eq.(\ref{Pk-1H}),
with the counterterm $\tW^2$ associated with mass and momentum conservation,
which ensures that $P_{\rm 1H}(k) \propto k^4$ at low $k$.
Again, this gives a ``halo-model'' power spectrum (\ref{Pk-halos}) that is identical
to Eq.(\ref{P-no-sc-1}) at all orders of perturbation theory. In particular, thanks to the
choice (\ref{S3-def}), it agrees with standard perturbation theory up to one-loop
order (and contains partial terms at all higher orders, generated through the
function $\varphi(y)$, as well as nonperturbative terms of the
form $e^{-1/\sigma^2}$).

\bibliography{ref1}   % name your BibTeX data base

\end{document}